\documentclass[usenatbib]{mnras}

\usepackage{graphicx}	
\usepackage{amsmath,amssymb,amsfonts,mathrsfs} 
\usepackage{bm}		
\usepackage{soul}
\usepackage[flushleft]{threeparttable}
\usepackage{multirow}
\usepackage{xspace}
\usepackage{Custom_Functions}

\defcitealias{gaikwad2016}{Paper-I}	
\defcitealias{puchwein2015}{P15}		
\defcitealias{haardt2012}{HM12}		
\defcitealias{khaire2015a}{KS15}		
\defcitealias{khaire2018a}{KS19}		
\defcitealias{gaikwad2018}{G18}	

\usepackage[usenames, dvipsnames]{color}
\definecolor{mycolor}{RGB}{0,128,0}
\definecolor{newaddcolor}{RGB}{255,0,255}

\usepackage[normalem]{ulem}
\newcommand{\PG}[1]{{\color{black} #1}}

\title[Effect of non-equilibrium ionization on IGM]{Effect of non-equilibrium ionization on derived physical conditions of the high-$z$ intergalactic medium}

\author[Gaikwad et.al]{Prakash Gaikwad$^{1,2,3}$\thanks{E-mail: \href{pgaikwad@ast.cam.ac.uk}{pgaikwad@ast.cam.ac.uk}}, 
Raghunathan Srianand$^{3}$,
Vikram Khaire$^{4}$ and 
\newauthor{Tirthankar Roy Choudhury$^{5}$}
\\
$^{1}$Institute of Astronomy, University of Cambridge, Madingley Road, Cambridge, CB3 0HA, UK \\
$^{2}$Kavli Institute for Cosmology, University of Cambridge, Madingley Road, Cambridge, CB3 0HA, UK \\
$^{3}$Inter-University Centre for Astronomy and Astrophysics (IUCAA), Post Bag 4, Pune 411007, India \\
$^{4}$Physics Department, Broida Hall, University of California Santa Barbara, CA 93106-9530, USA\\
$^{5}$National Centre for Radio Astrophysics, Tata Institute of Fundamental Research, Pune 411007, India}

\date{}

\pubyear{2015}

\begin{document}
\label{firstpage}
\pagerange{\pageref{firstpage}--\pageref{lastpage}}
\maketitle


\begin{abstract}
    Non-equilibrium ionization effects are important in cosmological hydrodynamical simulations but are computationally expensive.
    We study the effect of non-equilibrium ionization evolution and UV ionizing background (UVB)  
    generated with different quasar spectral energy distribution (SED) on the derived physical conditions of
    the intergalactic medium (IGM) at $2\leq z \leq 6$ using our post-processing tool ``Code for Ionization
    and Temperature Evolution'' (\citecode). 
    \citecode produces results matching well with self-consistent simulations more efficiently.
    The \HeII reionization progresses more rapidly in non-equilibrium model as compared to equilibrium models. 
    The redshift of \HeII reionization strongly depends on the quasar SED and occurs 
    earlier for UVB models with flatter quasar SEDs.
    During this epoch the normalization of temperature-density relation, $T_0(z)$, has a maximum while the slope, $\gamma(z)$, has a minimum, but occurring at different redshifts.
    The $T_0$ is higher in non-equilibrium models using UVB obtained with flatter quasar SEDs.
    \PG{While our models produce the observed median \HeII effective optical depth evolution
    and its scatter for equilibrium and non-equilibrium considerations, to explain the
observed cumulative distributions we may need to consider fluctuating UVB.}
    For a given UVB model, the redshift dependence of the \HI photo-ionization rate derived from the observed 
    \HI effective optical depth (\taueffHI) for the equilibrium model is different from that for the non-equilibrium model.
    This may lead to different requirements on the evolution of ionizing emissivities of sources. 
    We show that, in the absence of strong differential pressure smoothing effects, it is possible to recover the $T_0$ and $\gamma$ realised
    in non-equilibrium model from the equilibrium models generated by rescaling photo-heating rates while producing the same \taueffHI.
\end{abstract}


\section{Introduction}
\label{sec:introduction}

The \HI \lya forest absorption seen in the spectra of distant quasars in conjunction with
cosmological hydrodynamical simulations enable one to not only constrain cosmological parameters
\citep{weinberg1998,mcdonald2000,penton2000,phillips2001,mcdonald2005,viel2004a,shull2012b}
but also probe the thermal and ionization history of the intergalactic medium 
\citep[IGM,][]{schaye2000,faucher2008d,lidz2010,becker2011,becker2013}.
The outputs of a cosmological hydro-simulation depend on various ingredients such as (i) the assumed nature of
ionizing background, (ii) implementation of various feedback processes and (iii) assumptions involved
in the computations of thermal and ionization state of the gas. It is usually assumed that the
IGM is in ionization equilibrium with uniform ionizing background dominated by quasars and galaxies.
While most of these assumptions are valid for IGM in the post reionization era, they may not reflect the
reality when reionization is in progress or just completed \citep{puchwein2015,puchwein2018}.

\vspace{5mm}
We have enough observational evidences to suggest that the \HI reionization was completed around
$z \sim 5.5$ \citep{fan2001,fan2006,planck2014}. 
The \HeII \lya effective optical depth, measured towards rare UV bright quasars
using Hubble Space Telescope (HST), shows a strong evolution over $2.7\le z \le 4.0$ 
and a much smaller scatter at $z<2.7$ \citep{jakobsen1994,heap2000,kriss2001,shull2004,fechner2006,worseck2011,worseck2018}.
These observations are consistent with the IGM going through two major
reionization episodes with \HeII reionization being completed around $z\sim 2.7$, most likely driven by
luminous quasars. In such a scenario one expects a fresh influx of thermal energy and
entropy into the IGM at $2.7\le z \le 4.0$.
Interestingly, the nature of \HeII reionization can also be probed using well measured properties of \HI
\lya forest which can be easily accessed through ground based large telescopes. 
In particular one can use (i) the redshift evolution of temperature-density relation
in IGM which is driven by residual photo-heating from \HeII ionization that systematically broadens the \HI absorption lines \citep{schaye2000,theuns2002b,bolton2008,lidz2010,becker2011,
garzilli2012,boera2014,rudie2012,hiss2017,walther2018} and (ii) the presence or absence of structures in the redshift  evolution of
mean \HI \lya effective optical depth as a function of $z$ \citep{bernardi2003,faucher2008d,becker2013a}.

Interpretations of IGM observations will crucially depend on assumptions related to basic
ingredients of the cosmological simulations. In this work we investigate how the physical properties of the high-$z$
IGM is affected by (i) the allowed range in the \HeII photoionization rate (\GHeII)
for a given \HI photoionization rate (\GHI) originating from the uncertainties in UV spectral energy distributions
of quasars and (ii) the non-equilibrium effects in the calculations of ionization and thermal history of the gas.

The redshift evolution of cosmic ultra-violet radiation background (UVB) is computed by
solving a 1D radiative transfer of extreme UV photons emitted by sources (such as galaxies and quasars), and attenuated by 
the foreground gas in IGM \citep{haardt1996,faucher2008,haardt2012,khaire2015b,khaire2018a,puchwein2018}. 
The UV emissivity of quasars in the energy range relevant for \HeII reionization is highly uncertain due to uncertainties 
in mean spectral energy distribution of quasars \citep{khaire2017,khaire2018a}.
Since quasars are main sources of \HeII ionizing photons,
any uncertainty in its spectral energy distribution (specially in the extreme UV to soft X-ray range) can 
lead to uncertainties in the derived IGM parameters.

The thermal and ionization evolution of IGM in most of the hydrodynamical simulations 
are computed assuming the photo-ionization equilibrium. 
However, it takes some time for a parcel of gas, that got heated
recently by the ionization of \HeII,
to reach a new equilibrium. If the heating rate is faster than the time required for the gas
to reach the photo-ionization equilibrium then the gas parcel will remain in a non-equilibrium state
for a prolonged period of time. In such a cases it will be more appropriate to
consider non-equilibrium ionization evolution \citep[see][]{puchwein2015,puchwein2018}.
However, for a given UVB, the hydrodynamic simulation incorporating non-equilibrium ionization
evolution is expensive than a corresponding simulation performed using equilibrium consideration.
Here we are exploring an approach that will allow us to study the effect of UVB and non-equilibrium
evolution considering a wide range of parameter space in a computationally economical way.

In \citet{gaikwad2017a}, we have developed an efficient method to simulate the effect of UVB on ionization and thermal
evolution of the IGM in the post-processing step of \gtwo (i.e., hydrodynamical simulation without radiative processes) using a module
``\citefullform'' (\citecode). It was shown that most statistical distributions of IGM can be reproduced within 20 percent uncertainty
compared to self-consistent \gthree simulations when we consider signal-to-noise, spectral resolution and total
redshift pathlength typically achieved in
real observations.
In this work, we implement the non-equilibrium ionization evolution in \citecode (see \S \ref{sec:simulation}) 
and study its effect on \HeII reionization for a range of UVB models obtained by varying quasar spectral indices.

This paper is organised as follows. In section~\ref{sec:simulation}, we provide basic details of our simulations,
implementations of equilibrium and non-equilibrium calculations and various UVBs considered in this study.
In section~\ref{sec:results}, we provide detailed comparisons of our simulations with those from the literature,
study differences in the evolution of physical quantities in the non-equilibrium and equilibrium calculations and
discuss the implications of using equilibrium simulations to extract IGM parameters when the actual evolution
is governed by non-equilibrium processes. We summarize our results in section~\ref{sec:summary}.

Throughout this work we use flat $\Lambda$CDM cosmology with parameters 
$(\Omega_{\Lambda},\Omega_{m},\Omega_{b},\sigma_8,n_s, h,Y) = (0.69,0.31,0.0486,0.83,0.96,0.674,0.24)$ 
consistent with \citet{planck2014}.
All distances are given in comoving units unless specified.
\GHI expressed in units of $10^{-12} \: {\rm s}^{-1}$ is denoted by \GTW.

\section{Simulation}
\label{sec:simulation}

The high resolution simulations used in this work were run using the publicly available smooth particle hydrodynamics code \gtwo\footnote{http://wwwmpa.mpa-garching.mpg.de/gadget/} \citep{springel2005} with initial conditions generated at redshift $z = 99$ using the publicly available \twolpt\footnote{http://cosmo.nyu.edu/roman/2LPT/} code \citep{2lpt2012}. Our default simulation has a box size of $L_{\rm box} = 10 h^{-1}$ \cmpc, number of particles $N_{\rm particle} = 512^3$ and gas mass resolution, $m_{\rm gas} \sim 10^5 \: \Msun$.
We set gravitational softening length as 1/30$^{\rm th}$ of the inter particle distance.
We store the outputs from $z=6$ to $z=1.6$ in steps of $\Delta z = 0.1$ to track the density and temperature evolution of particles.

The publicly available version of \gtwo neither solves the ionization evolution equations nor incorporates the radiative heating or cooling. As a result, the temperature of low density unshocked gas (high density shocked gas) in \gtwo is lower (higher) than that expected in a self-consistent simulation. An updated version of \gtwo, known as \gthree, solves the ionization
and thermal evolution equations for a given 
UVB model self-consistently \citep{springel2005}. However, performing \gthree simulations for a wide range of UVB models is computationally expensive \citep[see, e.g., Table 6 in][]{gaikwad2018}. Also, the default version of \gthree solves the evolution equations assuming ionization equilibrium \citep[but see][where non-equilibrium ionization evolution has been incorporated within \gthree]{puchwein2015,puchwein2018}. 

To account for the above issues, we have developed a post-processing module called ``Code for Ionization and Temperature Evolution'' (\citecode) which can be used to incorporate the radiative heating and cooling such that one will be able to mimic various physical effects present in the simulations (such as pressure broadening) within few percent accuracy \citep{gaikwad2017a,gaikwad2018}. In \citet[][here after \citetalias{gaikwad2018}]{gaikwad2018}, we have shown that \citecode is flexible, efficient and can reproduce results consistent with self-consistent \gthree simulations for equilibrium ionization evolution. Further, it also allows us to model the non-equilibrium ionization and thermal evolution. It has been shown in \citetalias{gaikwad2018} that the most accurate results (when compared to a full \gthree runs) are obtained when \gtwo is run with elevated temperature floor of $10^4$ K, motivated by the fact that the typical temperature of photo-ionized IGM is few times $10^4$ K \citep[at overdensity of $\Delta = 1$,][]{hui1997}.

In the redshift range of our interest ($2 \leq z \leq 6$), the \HeII ionizing UVB can
have large spatial fluctuations 
\citep{furlanetto2008a,dixon2014,leplante2016,leplante2017a,leplante2017b, davies2017}
as one is most likely probing the end stages of \HeII reionization.
These fluctuations arise mainly because the mean free path of \HeII ionizing 
photons is smaller than the mean separation between the quasars. 
However, in this work we use UVB from synthesis models which implicitly assume 
that the UVB is spatially uniform (but is evolving with time). 
This, combined with the fact that our simulation box is too small to capture properties of
individual \HeIII region around a typical bright quasar, 
implies that we do not account for the patchy \HeII reionization process. 
We assume instead that the filling fraction of \HeIII regions and the \fHeIII in our 
simulation box are equivalent \citep[][here after \citetalias{puchwein2015}]{puchwein2015}.
Furthermore, the heating rate during \HeII reionization 
depends not only on the nature of UVB but also 
on the speed of the ionization front in IGM \citep{daloisio2018}. 
On the other hand, our simulation can resolve \HI \lya forest at the thermal broadening scales
and has the resolution consistent with that has been typically used in the echelle spectroscopy
for studying the \lya forest.

\subsection{Temperature and Ionization Evolution Equation}
\label{subsec:temperature-evolution}
The thermal evolution of IGM is governed by following equation \citep{hui1997,gaikwad2017a,gaikwad2018},

\begin{eqnarray}\label{eq:temperature-evolution}
    \frac{dT}{dt} &=& -2HT + \frac{2T}{3\Delta} \: \frac{d \Delta}{dt} +  \frac{dT_{shock}}{dt} + \frac{T}{\sum \limits_{X} f_{\rm X}} \: \frac{d \sum \limits_{\rm X} f_{\rm X}}{dt}
\nonumber\\
&+&  \frac{2}{3 \: k_B \: n_b} \; (\mathscr{H} - \mathscr{C}) 
\end{eqnarray}
where $H$, $k_B$, $\Delta$ and $n_b$ are Hubble parameter, Boltzmann constant, overdensity and number density of baryons respectively.
    The symbol $f_{\rm X} \equiv n_{\rm X} / n_{\rm H}$ where $n_{\rm X}$ and $n_H$ are number densities of species
$X \in [$\HI,\HII, \HeI, \HeII, \HeIII, $e]$ and  hydrogen respectively.
$\mathscr{H}$ and $\mathscr{C}$ are the total heating and cooling rates per unit volume (${\rm ergs \; cm^{-3} \; s^{-1}}$) respectively.
The five terms on the right hand side of the above equation, respectively, represent rate of change of temperature due to,
(i) Hubble expansion,
(ii) adiabatic heating or cooling, 
(iii) shock heating from structure formation, 
(iv) change in internal energy and
(v) radiative heating or cooling processes.
The first three terms on the right hand side of the above equation are computed self-consistently in \gtwo.
The last two terms are calculated in \citecode for each particle by solving the following set of ionization 
evolution equations assuming primordial composition (H and He only) of gas,

\begin{equation} \label{eq:ionization-evolution-equation}
\begin{aligned}
    \alpha_{\text{HII}} \: n_e \: f_{\rm HII} - f_{\rm HI} \: (\Gamma^{\gamma}_{{\rm H\textsc{i}}} + \Gamma^{\rm e}_{{\rm H\textsc{i}}} \: n_e) &= \frac{d f_{\rm HI}}{dt} \\
    \alpha_{\text{HeII}} \: n_e \: f_{\rm HeII} - f_{\rm HeI} \: (\Gamma^{\gamma}_{{\rm He\textsc{i}}} + \Gamma^{\rm e}_{{\rm He\textsc{i}}} \: n_e) &= \frac{d f_{\rm HeI}}{dt} \\ 
- \alpha_{\text{HeIII}} \: n_e \: f_{\rm HeIII} + f_{\rm HeII} \: (\Gamma^{\gamma}_{{\rm He\textsc{ii}}} + \Gamma^{\rm e}_{{\rm He\textsc{ii}}} \: n_e) &= \frac{d f_{\rm HeIII}}{dt} \\
    f_{\rm HI} +  f_{\rm HII} &= 1 \\  
    f_{\rm HeI} + f_{\rm HeII} + f_{\rm HeIII} &= y \\
    f_{\rm HII} + f_{\rm HeII} + 2 f_{\rm HeIII} &= f_{\rm e} \\
\end{aligned}
\end{equation}
where $\Gamma^{\gamma}_{{X_i}}$, $\Gamma^{\rm e}_{{X_i}}$, $\alpha_{X_i}(T)$ and $n_{\rm e}$
are photoionization rate, collisional ionization rate, recombination rate coefficient and number density of electrons respectively. 
The fraction $y = Y m_{\rm H} / [m_{\rm He} \: (1-Y)]$ is He abundance by number where $m_{\rm H}, 
m_{\rm He}$ are the atomic mass of H, He and $Y$ is He abundance by mass respectively. 
We use collisional ionization and recombination rates from \citet{theuns1998b}.
The term $\mathscr{C}$ and $\mathscr{H}$ can be calculated by summing the cooling rates \citep[][see their Table B1]{katz1996,theuns1998b} and heating rates
as given by, 

\begin{eqnarray}\label{eq:heating-cooling-rate}
    \mathscr{H} &=& \; (f_{\rm HI} \: \epsilon_{\gamma {\rm HI}} \; + \; f_{\rm HeI} \: \epsilon_{\gamma {\rm HeI}}  \; + \; f_{\rm HeII} \: \epsilon_{\gamma {\rm HeII}}) / n_{\rm H}
\nonumber \\  
\mathscr{C} &=& \sum\limits_{k=1}^{11} \; c_k(T, z, {X_i}) 
\end{eqnarray}
where $\epsilon_{\gamma {\rm X}}$ is photo-heating rates for species $X \equiv [\HIeqn, \HeIeqn, \HeIIeqn]$ 
and $c_k(T_,z, {X_i})$ is cooling rate coefficient for all the relevant cooling processes such as 
collisional ionization, recombination, dielectronic recombination,
collisional excitation, Bremsstrahlung and inverse Compton cooling. 
We use $c_k(T_,z, {X_i})$ consistent with that from \citet[][see Table B1]{theuns1998b}. 
The photoionization and photo-heating rates are obtained from the assumed uniform UVB model. 
We also use the rate coefficients as given in \citet{lukic2015} and find that 
the maximum difference in the results are less than 8 percent.

\subsection{Equilibrium and Non-equilibrium Ionization Evolution}
\label{subsec:eqbm-non-eqbm-ionization}
In the simulation runs using the default version of \gthree, Eq. \ref{eq:ionization-evolution-equation} is solved for equilibrium ionization 
conditions i.e., setting right hand side of the first three relations in 
Eq. \ref{eq:ionization-evolution-equation} to zero 
and excluding the explicit time dependence of ionization fraction \citep[exceptions being][]{puchwein2015,puchwein2018}. 
In our case, to solve equilibrium ionization evolution equations (at any $z$) numerically,
we start with an initial guess value for  electron number density ($n_e \sim n_{H}$). 
We then solve the set of simultaneous equations (first $5$ equations in Eq. \ref{eq:ionization-evolution-equation})
to compute the ionization fraction for different states of H and He.
We use the last relation in Eq. \ref{eq:ionization-evolution-equation} to update the value of electron number 
density $n_e$.
The recombination rates ($\alpha_{\rm X}$) depend on the temperature
and we use the updated temperatures at each time step to compute the ionization fractions accurately.
We repeat the process with updated value of $n_e$ until the values of all the 
variables converge to an absolute tolerance of $10^{-7}$.

Using equilibrium solutions, however, may not be a good approximation if the time scales of radiative processes (recombination or photo-ionization) become longer than the Hubble time or the time scale over which physical quantities like gas temperature evolve.
In particular, such an approximation is not expected to be valid during \HeII (or \HI) reionization since it takes some time
for an ionized parcel of gas to reach a new ionization equilibrium.

The non-equilibrium solutions to Eq. \ref{eq:ionization-evolution-equation} is non-trivial as the 
explicit time dependent equations (first three relations in Eq. \ref{eq:ionization-evolution-equation}) 
tend to be stiff owing to the different time scales involved in the problem. 
We solve the non-equilibrium ionization evolution equation (Eq. \ref{eq:ionization-evolution-equation}) 
using Sundials CVODE library (\textit{python} version) \citep{cohen1996,oppenheimer2013,puchwein2015}.
This library integrates the set of stiff ordinary differential equations with variable-order, 
variable-step using Backward Differentiation Formula  (BDF) methods.
We use the relative error tolerance of $10^{-7}$ for various ionization fractions of H and He in CVODE library.

It is important to note that the equilibrium or non-equilibrium models
discussed in this work refers to the nature of the ionization evolution of the gas in IGM.
In both models, we solve temperature evolution equation with explicit time dependence that  is
we account for  $dT/dt$ term on left hand side of Eq. \ref{eq:temperature-evolution}.

The difference between equilibrium and non-equilibrium ionization evolution of \HeII 
could affect the changes in electron number density and thermal history of the IGM.
As the contribution of \HeII to the number density of electrons is small ($n_{\rm H} / n_{\rm He} \sim 12$),
change in $n_e$ due to reionization of \HeII does not influence the ionization state of \HI significantly.
However, the effect on the thermal history of the IGM could be significant depending 
on whether one uses equilibrium and non-equilibrium ionization evolution. 
Since the recombination rate depends on the temperature of the IGM, the ionization state of \HI can be affected significantly (see Eq. \ref{eq:ionization-evolution-equation}).
In addition, the pressure smoothing effects are also expected to be significantly different between equilibrium and non-equilibrium ionization evolution.
Keeping this in mind, our main aim in this work is to explore the effect of equilibrium and non-equilibrium ionization evolution 
(i) on thermal history of IGM,
(ii) on the derived \HI photo-ionization rate \GHI at $2 < z < 4$ and 
(iii) on the redshift evolution of \HeII effective optical depth.
We will discuss these issues in detail in \S \ref{sec:results}.

\subsection{UVB Models}
\label{subsec:uvb-models}
The main inputs for the \citecode and our non-equilibrium calculations are the photoheating 
and photoionization rates obtained from the UVB synthesis models. 
In such synthesis models, the UVB at any redshift $z_0$ is obtained by solving 
the radiative transfer of extreme UV photons emitted by sources (quasars and galaxies) and filtered by 
the gas in IGM at all $z>z_0$ \citep[see e.g][]{haardt1996,faucher2008,haardt2012,khaire2015b,khaire2018a,puchwein2018}. 
The UV emissivities of sources are obtained by using the observed luminosity functions at a single
rest wavelength combined with 
their mean spectral energy distribution (SED). 
However, these emissivity estimates are highly uncertain because 
(i) there are very little observational constraints on either ionizing emissivities or the SED of galaxies in the relevant energy ranges, 
(ii) there is no consensus on the ionizing SED of quasars \citep[see table 1 of][]{khaire2017}, and
(iii) the faint end slopes and the limiting magnitudes of quasars and galaxies,  to be used
while integrating the luminosity functions to obtain emissivities are 
not well constrained at $z>3$ \citep{giallongo2015,khaire2015a,khaire2016,kulkarni2018}. 
To partially circumvent the emissivity uncertainties, the UVB synthesis models usually 
tune the unconstrained ionizing emissivity from galaxies in order to match with the observed 
\HI photoionization rates.
This is performed by changing escape fraction (\fesc) of \HI ionizing photons from galaxies 
under the assumption that the SED of galaxies is determined by stellar population synthesis 
models and extinction by dust. It is also usually assumed that   
galaxies with typical stellar populations (i.e population II) do 
not contribute significantly to high energy photons that can ionize \HeII \citep{dayal2018}. 

Unfortunately, fixing the \HeII ionizing emissivity of quasars for $E > 54.4$ eV is also
difficult as  quasar SEDs are not measured in these energy ranges and
constraints on 
\HeII photoionization
rates obtained from high-z IGM are just beginning to emerge
\citep[for e.g.,][]{khaire2017,worseck2018}. 
To overcome this issue in \HeII ionizing UVB,
we use the allowed range of UVB models presented in  \citet[][here after \citetalias{khaire2018a}]{khaire2018a}
obtained with different \HeII ionizing emissivities. 
In \citetalias{khaire2018a}, the quasar emissivity at non-ionizing wavelengths (Far UV) is taken from the updated estimates by 
\citet{khaire2015b} and the ionizing emissivity is obtained by varying quasar 
SED using allowed range of power law indices. The measurements of quasar SEDs are obtained by fitting power 
law relation ($f_{\nu} \propto \nu^{-\alpha}$) to the stacked mean ionizing spectrum of quasars (at $E > 13.6$ eV). 
The recent measurements of such power law index $\alpha$ has been reported to have values from $0.7$ to $2.5$ 
\citep{shull2012b, stevans2014, lusso2015, tilton2016, lusso2018}. 
In this work, we have used four UVB  models calculated using $\alpha$ in the range of $1.4$ to $2.0$ with the interval of $0.2$ 
\citep{khaire2018a}. 
In addition to these four UVB models we also use \citet[][here after \citetalias{haardt2012} that used $\alpha=1.57$]{haardt2012} UVB for comparison with results of \citetalias{puchwein2015}.

In order to facilitate easy identification of models, we refer to them using nomenclatures as identified in first column of Table \ref{tab:model-details}. 
Model NE-KS19-2.0 refers to non-equilibrium ionization evolution of gas exposed to 
\citetalias{khaire2018a} UVB with quasar SED obtained using a power law index of $\alpha = 2.0$. 
The model EQ-KS19-1.4 corresponds to the equilibrium ionization evolution for \citetalias{khaire2018a} UVB with quasar SED having $\alpha = 1.4$.
All models listed in Table \ref{tab:model-details} are obtained by post-processing \gtwo simulation 
($L_{\rm box} = 10 h^{-1}$ \cmpc, $N_{\rm particle} = 512^3$, see \S \ref{sec:simulation}) with \citecode.

\begin{table*}
\caption{Details of models studied in this work. Same for additional models discussed in this work are given in the Appendix \ref{app:additional-simulations}}
\begin{threeparttable}
\centering
\begin{tabular}{ccccccccc}
\hline \hline
Model Name\tnote{a} & Code\tnote{b} & Ionization evolution & UVB & Quasar spectral index ($\alpha$)\tnote{c} & Reference \\
\hline \hline
EQ-KS19-1.4 & \gtwo + \citecode  & equilibrium      & \citetalias{khaire2018a}  & 1.4  & This work \\ 
NE-KS19-1.4 & \gtwo + \citecode  & Non-equilibrium  & \citetalias{khaire2018a}  & 1.4  & This work \\ 
EQ-KS19-2.0 & \gtwo + \citecode  & equilibrium      & \citetalias{khaire2018a}  & 2.0  & This work \\ 
NE-KS19-2.0 & \gtwo + \citecode  & Non-equilibrium  & \citetalias{khaire2018a}  & 2.0  & This work \\ 
EQ-HM12     & \gtwo + \citecode  & equilibrium      & \citetalias{haardt2012}   & 1.57 & This work \\ 
NE-HM12     & \gtwo + \citecode  & Non-equilibrium  & \citetalias{haardt2012}   & 1.57 & This work \\ 
EQ-HM12-G18 & \gthree            & equilibrium      & \citetalias{haardt2012}   & 1.57 & \citet{gaikwad2018} \\ 
NE-HM12-P15\tnote{d} & \gthree   & Non-equilibrium  & \citetalias{haardt2012}   & 1.57 & \citet{puchwein2015} \\ 
\hline \hline
\end{tabular}
\begin{tablenotes}
\item[a] In the main text of this paper, we present the results of total $4$ different models. 
         We refer readers to appendix for additional $16$ models analysed in this work.
         In the appendix, these models are referred with prefix L10-N512-G2 (L10-N512 model performed using \gtwo + \citecode)
         attached to the model name. For example, the model EQ-KS19-1.4 is referred as L10-N512-G2-EQ-KS19-1.4.
         All simulation outputs were store from $z=6$ to $z=1.6$ in steps of $\Delta z = 0.1$ (except NE-HM12-P15).
\item[b] \gtwo simulation is performed with $L_{\rm box} = 10 h^{-1}$ \cmpc, $N_{\rm particle} = 2 \times 512^3$ 
         (L10-N512) and gas mass resolution $\delta m \sim 10^5 \: \Msun$.
         Post-processing module \citecode is applied on \gtwo output. 
         \citecode provides flexibility to change UVB and perform equilibrium / non-equilibrium ionization evolution efficiently. 
\item[c] The UVB models are generated by varying the uncertain quasar spectral slopes $\alpha$, defines as $f_{\nu} \propto \nu^{-\alpha}$ \citetalias{khaire2018a}. 
         We vary $\alpha$ from 1.4 to 2.0 in steps of 0.2. 
\item[d] NE-HM12-P15 simulations are performed with $L_{\rm box} = 20 h^{-1}$ \cmpc, $N_{\rm particle} = 2 \times 512^3$
         \citep[see, ][]{puchwein2015}.
         
\end{tablenotes}
\end{threeparttable}
\label{tab:model-details}
\end{table*}

\section{Result}
\label{sec:results}

\subsection{Comparison between equilibrium and non-equilibrium models}

We first present our results on how the IGM properties depend on whether
we are solving the Eq.~\ref{eq:ionization-evolution-equation}
using the equilibrium approximation or taking into account the explicit time evolution.
We also demonstrate how well our post-processing approach (\citecode) captures
all the basic results seen in self-consistent simulations.

\subsubsection{Evolution of \HI and \HeII fraction}
\label{subsec:fHI-fHeII-evolution}
The evolution of \HI and \HeII fractions are important for understanding the effect
of UVB on ionization state of the IGM. 
Furthermore, the excess kinetic energy carried away 
by electrons is responsible for changing 
the thermal state of IGM during reionization. 
The amount of photo-heating is directly proportional to the \HI and \HeII fractions 
(see Eq. \ref{eq:heating-cooling-rate}).
We define a volume average \HI and \HeII fraction in following way \citep{puchwein2015}
\begin{equation} \label{eq:ionization-fraction}
    f_{\rm HI} = \frac{n_{\rm HI}}{n_{\rm H}} \;\;\;{\rm and}\;\;\; f_{\rm HeII} = \frac{n_{\rm HeII}}{(n_{\rm HeI} + n_{\rm HeII} + n_{\rm HeIII})}
\end{equation}
where $n_{\rm X}$ is volume average number density of a species X.

\InputFigCombine{Fraction_Evolution_Literature_2.pdf}{160}{%
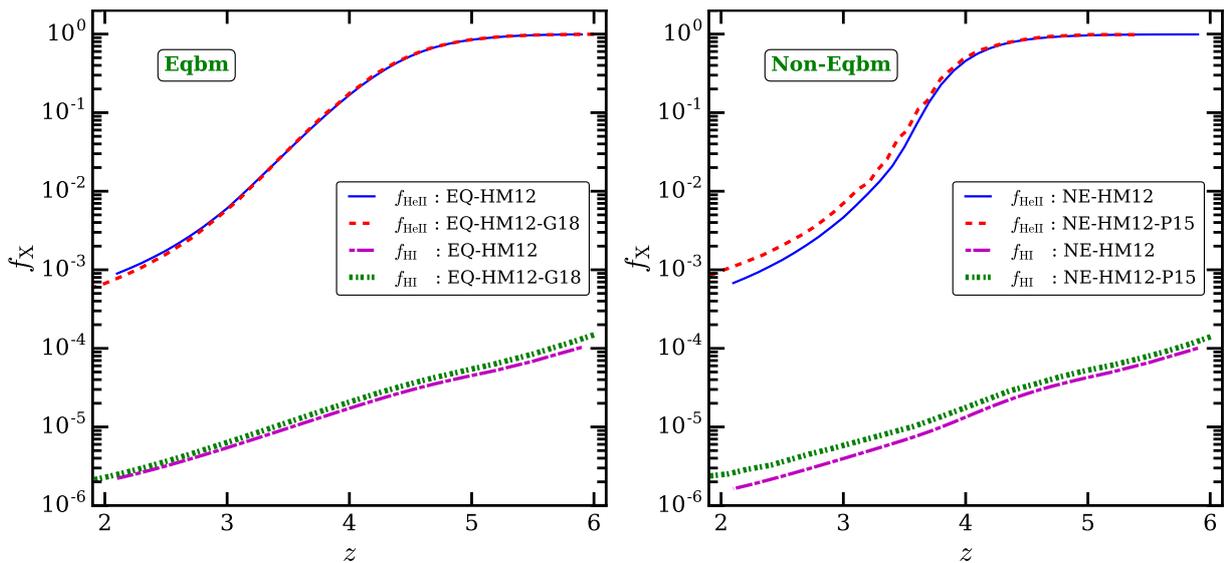
    Left panel shows the comparison of volume averaged \fHI and \fHeII evolution
    for equilibrium model from this work (EQ-HM12, \fHI in magenta dashed-dotted line, \fHeII in blue solid line) 
    with that from self-consistent models derived in 
    \citetalias{gaikwad2018} (EQ-HM12-G18, \fHI in green dotted line, \fHeII in red dashed line).
    The \fHI and \fHeII evolution in two models are in good agreement.
    Right panel is similar to left panel except the comparison is shown for non-equilibrium
    models where we compare the \fHI and \fHeII evolution in this work with that from 
    \citetalias{puchwein2015} (\fHI in green dotted line, \fHeII in red dashed line).
    The small mismatch between the two models is mostly due to the differences in the rate 
    coefficients and cosmological parameters (simulation box size and number of particles are same).
    All the curves are shown for \citet{haardt2012} UVB model.  
}{\label{fig:consistency-ionization-fraction}}

We show the \fHeII and \fHI evolution for equilibrium (left panel) 
and non-equilibrium (right panel) models using \gtwo + \citecode with HM12 UVB in Fig. \ref{fig:consistency-ionization-fraction}. We also show the same two quantities obtained using an equilibrium \gthree simulation (EQ-HM12-G18), identical to that used in \citetalias{gaikwad2018}, in the left panel of Fig. \ref{fig:consistency-ionization-fraction}. Finally, we compare the \fHeII and \fHI evolution from our calculations with
those obtained from the \gthree simulations of \citetalias{puchwein2015} (NE-HM12-P15) in the right panel in Fig. \ref{fig:consistency-ionization-fraction}. 
We remind the readers that the \fHeII and \fHI evolution in NE-HM12-P15 model is calculated self-consistently by
solving the non-equilibrium ionization evolution equations in the \gthree code.

From the left panel of the figure, we find that ion fractions obtained using our implementation of the equilibrium model matches within 14 percent with that of
the \gthree simulations. Similarly, we find a good agreement between results of our non-equilibrium model (NE-HM12) and the models of \citetalias{puchwein2015} (NE-HM12-P15) as can be seen from the right panel of the figure.
The difference between the two non-equilibrium models is within $\sim 20$ percent and major contribution to this can be attributed to the 
differences in the rate coefficients, cosmology and non-equilibrium solver used 
in this work and \citetalias{puchwein2015}\footnote{The 
difference due to rate coefficient is less than 8 percent \citep{lukic2015}.}.
Note that \fHeII is the main quantity governing the 
ionization and thermal evolution of IGM at $2 \leq z \leq 6$. 
Since \fHeII evolution for equilibrium and non-equilibrium models obtained using \gtwo + \citecode match with those obtained from full \gthree runs, the thermal evolution in these models are also consistent with each other. 
The consistency of the thermal evolution for equilibrium and non-equilibrium models with that from \gthree 
\citepalias[][for non-equilibrium model]{puchwein2015}
has already been shown in Fig. 4  of \citetalias{gaikwad2018}.

Thus, Fig. \ref{fig:consistency-ionization-fraction} validates our implementation of 
equilibrium and non-equilibrium ionization evolution in \citecode.
In particular, the agreement of \fHeII evolution at 20 percent
level between \citetalias{puchwein2015} and our implementation
is important for modelling the IGM at $2 \leq z \leq 6$. As we will show later
the differences in \fHeII evolution between equilibrium and non-equilibrium calculations
are much higher than the difference we find between exact calculations of non-equilibrium
\fHeII and our calculations using \citecode.

The left (right) panel of Fig. \ref{fig:ionization-fraction-combine} shows the evolution of \fHeII (\fHI) for two different \citetalias{khaire2018a}
UVB models with $\alpha = 1.4$ and $2.0$ respectively\footnote{We have also done similar analysis (i.e., 
    evolution of $f_{\rm HI}, \: f_{\rm HeII}, \: T_0$ and $\gamma$) for $\alpha=1.6$ and $1.8$.
However for simplicity, we show the results only for $\alpha=1.4$ and $2.0$ UVB models.}. 
From the left panel, we see that \fHeII $\sim 1$ at high redshifts and decreases to $\sim 10^{-2} - 10^{-3}$ around $z \sim 3$.
As expected, the exact redshift of \HeII reionization (for example say when \fHeII $\le 0.025$) depends on our choice of $\alpha$;
higher values of $\alpha$ provide lower \HeII ionizing emissivity delaying the redshift of \HeII reionization.
This result is qualitatively similar to \fHeII evolution obtained in \citet{puchwein2018}.
In this analysis, we assume that $\alpha$ is not evolving with redshift.
Any redshift dependence of $\alpha$ will complicate the redshift dependence of ion fractions and temperature.

Concentrating on the $\alpha=1.4$ case, we find that \fHeII is consistently smaller for equilibrium model (EQ-KS19-1.4) than that for non-equilibrium model (NE-KS19-1.4) during \HeII reionization at $3.7 < z < 5$. This is because \fHeII in the equilibrium model is equated to the recombinations at each time step \citep{puchwein2015}.
Since recombinations are proportional to $\Delta^{0.6}$ (see Eq. \ref{eq:fHeII-Delta-relation}), \fHeII in equilibrium models is decided 
by the instantaneous density of particles.
On the other hand, \fHeII in non-equilibrium model is decided by 
the \fHeII in previous time step, 
\HeII photo-ionization rate ($\sim \Gamma^{-1}_{\rm HeII}$), 
the recombination time scale
and the instantaneous density of gas. The photo-ionization and recombination time
scales evolve with redshift due to the \GHeII and thermal evolution. 
Irrespective of our choice of $\alpha$,\fHeII is found to be larger in 
non-equilibrium models compared to that of equilibrium models prior to the 
epoch of \HeII reionization.
This is probably related to a steep rise in gas temperature we see in the non-equilibrium models
within a typical recombination time-scale.
Once the \fHeII becomes small enough (\fHeII $\leq$ 0.025), we find that the ionization evolution is governed by the balance between
photo-ionization and recombination. In the case of non-equilibrium models, having slightly
higher temperatures, should result in lower \fHeII than that for equilibrium model. This is what
we see in Fig.~\ref{fig:ionization-fraction-combine}.
This trend seen in \HeII for post reionization epoch  is also
seen in the \fHI evolution (right panel in Fig.~\ref{fig:ionization-fraction-combine}).
The redshift  at which the photo-ionization equilibrium is reached is different for different UVB models.
For example, for $\alpha = 1.4$ and $2.0$ the redshifts are $z=3.7$ and $3.2$ respectively. 
Below these redshifts, the \fHeII fraction for non-equilibrium is consistently smaller than that for equilibrium models.

\begin{table}
\centering
\caption{Extent of \HeII reionization}
\begin{threeparttable}
\begin{tabular}{cccc}
\hline \hline
Model                   & $z_{\rm start}$\tnote{a} & $z_{\rm end}$\tnote{b} & $\Delta z$\tnote{c} \\ 
\hline
EQ-KS19-1.4 & 5.03 & 3.70 & 1.33 \\ 
NE-KS19-1.4 & 4.60 & 3.70 & 0.90  \\
EQ-KS19-2.0 & 4.50 & 3.20 & 1.30  \\
NE-KS19-2.0 & 4.05 & 3.20 & 0.85 \\
\hline \hline
\end{tabular}
\begin{tablenotes}
\item[a] $z_{\rm start}$ corresponds to the epoch when $f_{\rm HeII} = 0.9$ see Fig.\ref{fig:ionization-fraction-combine}
\item[b] $z_{\rm end}$ corresponds to the epoch when $f_{\rm HeII} = 0.025$ see Fig.\ref{fig:ionization-fraction-combine}
\item[c] $\Delta z = z_{\rm start} - z_{\rm end}$ represent extent of \HeII 
         reionization.
\end{tablenotes}
\end{threeparttable}
\label{tab:HeII-reionization-extent}
\end{table}

The extent of \HeII reionization is different when non-equilibrium effects are taken into account. 
To quantify the differences in the extent of \HeII reionization, 
we assume that the \HeII reionization starts at redshift $z_{\rm start}$ corresponding to 
the epoch when $f_{\rm HeII} = 0.9$ and it ends at redshift $z_{\rm end}$ when \fHeII drops below $0.025$.
This choice of $z_{\rm end}$ is motivated by the fact that equilibrium and 
non-equilibrium \fHeII cross each other at $f_{\rm HeII} = 0.025$ irrespective of the UVB model we use.
Table \ref{tab:HeII-reionization-extent} summarizes the $z_{\rm start}$, $z_{\rm end}$ and redshift interval ($\Delta z$) for \HeII reionization for 
the UVB models shown in Fig. \ref{fig:ionization-fraction-combine}.
The $z_{\rm start}$ is consistently lower for non-equilibrium models as compared to that from
equilibrium model. As a result, the \HeII reionization in equilibrium case  is more extended with $\Delta z \sim 1.3$. As we discussed earlier, the $z_{\rm start}$  and $z_{\rm end}$ are smaller for $\alpha = 2.0$ than for $\alpha = 1.4$ because of lower \HeII ionizing emissivity in the former case.

The right panel in Fig.\ref{fig:ionization-fraction-combine} shows evolution of \fHI for models with $\alpha = 1.4$ and 2.0.
Even though non-equilibrium ionization evolution effects are important during \HeII
reionization ($2 \leq z \leq 4$), photo-ionization equilibrium for \HI is still a good approximation as \HI is highly
ionized. The recombination rate required to maintain this highly ionized state can be achieved over a short time-scale. 
Therefore, at $z < 5$, \fHI is consistently smaller for non-equilibrium models than that for equilibrium model.
The variation in \fHI seen at $z < 5$ is consistent with evolution of \fHeII and temperature of IGM \citep[also see][]{puchwein2015}.
The temperature of IGM is consistently larger for non-equilibrium models than that for
corresponding equilibrium models at $z < 5$ (see \S \ref{subsec:T0-gamma-evolution} for details). 
Since $f_{\rm HI} \propto T^{-0.7}$ for photo-ionization equilibrium, the \fHI 
obtained in non-equilibrium model is consistently smaller.

\InputFigCombine{Fraction_Evolution_Combine.pdf}{175}{%
    Comparison of volume averaged \fHeII (left panel) and \fHI (right panel) evolution from \citetalias{khaire2018a} for equilibrium
    and non-equilibrium ionization evolution. 
    The evolutions of \fHI and \fHeII are shown for two \citetalias{khaire2018a} UVBs compiled for two different
    quasar spectral indices ($\alpha = 1.4$ and $2.0$). 
    Comparison of EQ-KS19-2.0 (magenta diamonds) with NE-KS19-2.0 (green stars) in left panel 
    shows that at $z>3.2$ \fHeII is under predicted in equilibrium case than that in non-equilibrium case.
    On the other hand at $z<3.2$, \fHeII is lower in equilibrium case than that from non-equilibrium due to larger $T_0$.
    Comparison of NE-KS19-1.4 (blue squares) with NE-KS19-2.0 (green stars) shows  that
    \fHeII is larger for larger $\alpha$ (this is valid for equilibrium case also) indicating that redshift of \HeII reionization strongly depends on $\alpha$. 
    Right panel shows that for a given UVB model, evolution of \fHI is similar for equilibrium and non-equilibrium case at $z \geq 4.5$.
    However at $z < 4.5$, \fHI in the equilibrium case is consistently more than that for non-equilibrium case for a given UVB model. 
    This is because $T_0$ (and \fHeII) in non-equilibrium model is systematically larger than that in equilibrium model at $z < 4.5$ (See Fig. \ref{fig:T0-gamma-evolution-KS19}).
}{\label{fig:ionization-fraction-combine}}
Another interesting effect to note from Fig.~\ref{fig:ionization-fraction-combine} is the appearance of
a kink like feature at $z\sim 3.8$ in the \fHI curve for EQ-KS19-1.4 model. Such a kink is not apparent in
other cases. As we will see latter this is related to a sharp increase in temperature with $z$ seen in
 EQ-KS19-1.4 model. We come back to this in more detail in following sections.

\subsubsection{Temperature-Density Relation}
\label{subsec:TDR}
\InputFigCombine{Particle_EoS.pdf}{175}{%
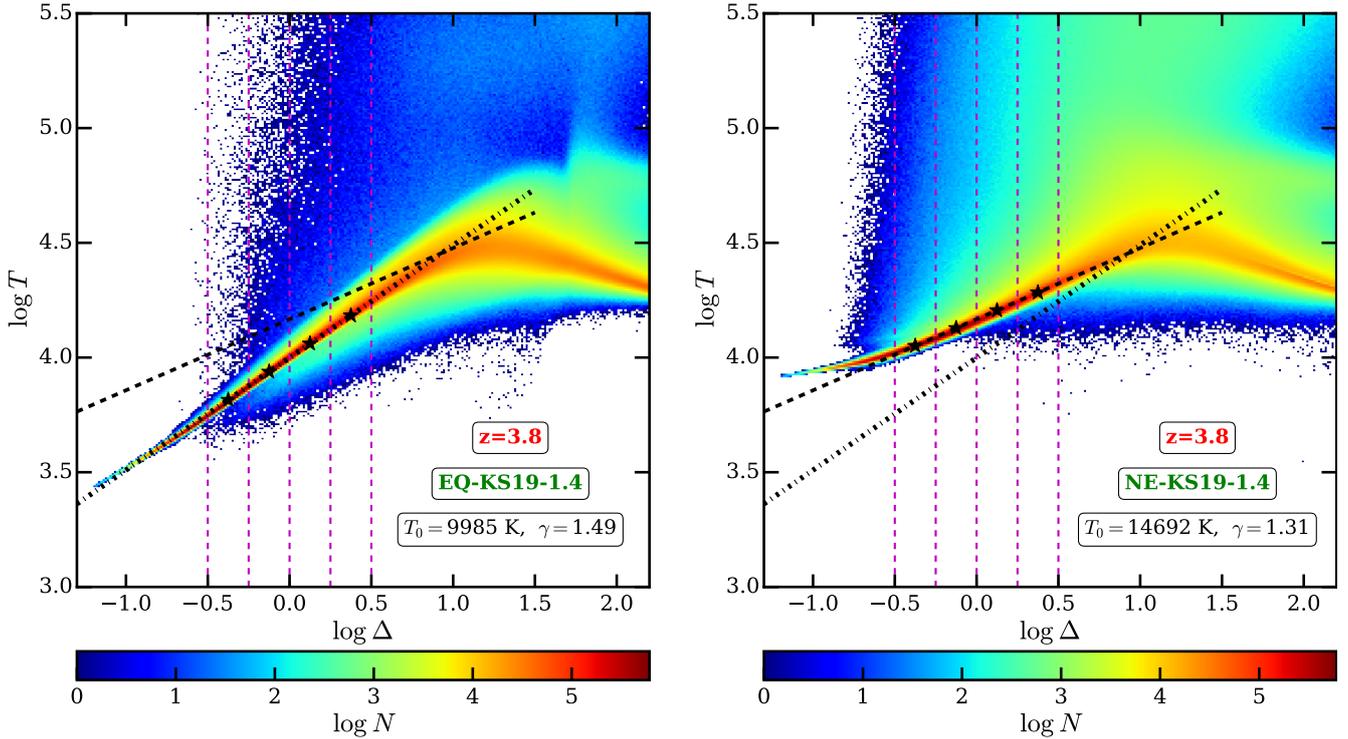
    Left and right panels show the TDR for EQ-KS19-1.4 and NE-KS19-1.4 models respectively at $z=3.8$
    (similar to Fig. 2 in \citetalias{puchwein2015}).
    The color scheme shows the density of points in logarithmic scale.
    The power-law is fitted to TDR by calculating the median temperature (black stars) 
    in $4$ $\log \Delta$ bins (shown by magenta dashed vertical lines).
    In both panels, the fitted power laws for equilibrium and non-equilibrium model 
    are shown by dotted and dashed black lines respectively.
    The TDR is flatter ($\gamma$ is smaller) and has higher normalization ($T_0$) in non-equilibrium model as compared to equilibrium model.
    At $\log \Delta < -0.5$, the TDR deviates from power-law for non-equilibrium models.
    In each panel we also provide the best fit values of $T_0$ and $\gamma$.
    Our method of calculating $T_0$ and $\gamma$ is different than that 
    from \citetalias{puchwein2015} (see \S \ref{subsec:TDR} for details).
}{\label{fig:TDR}}

The low density IGM follows a power-law temperature-density relation (TDR) of the form $T=T_0 \: \Delta^{\gamma-1}$
where $T_0$ and $\gamma$ are temperature of IGM at cosmic mean density ($\Delta=1$) and
slope of TDR respectively \citep{hui1997}.
For a given simulation at redshift $z$, we calculate $T_0$ and $\gamma $ as follows,
\begin{enumerate}
    \item We collect the temperature of all the particles in different $\log \Delta$ bins centred 
          at $[-0.375, -0.125, 0.125$ and $0.375]$ having width of $0.25$. 
          The choice of these bins is motivated by the fact that the TDR deviates from power-law at $\log \Delta < -0.5$ for non-equilibrium models (see Fig. \ref{fig:TDR})
          whereas the scatter in temperature is more at $\log \Delta > 0.5$ for equilibrium and non-equilibrium models.
    \item We then calculate the median temperature ($\log T$) of particles in each $\log \Delta$ bin.
    \item Finally, we obtain best fit $T_0$ and $\gamma$ by fitting straight line,  $\log T = \log T_0 + (\gamma-1)~\log~\Delta$, to the median $\log T$ measured in each  $\log \Delta$ bin.
\end{enumerate}

Fig. \ref{fig:TDR} shows the comparison of TDR from the equilibrium (left
panel) and non-equilibrium (right panel) models at $z=3.8$ for the KS19-1.4
UVB.  As expected, most of the points follow a power law TDR in equilibrium and
non-equilibrium cases.  However at $\log \Delta < -0.5$, the TDR in
non-equilibrium model is flattened.
This deviation at $\log \Delta \leq -0.5$ from a single power law for
non-equilibrium model occurs because \fHeII in the non-equilibrium case does
not only depend on the instantaneous density of particles (see Appendix
\ref{app:semi-numerical} for details).  As a result, \fHeII at $\log \Delta <
-0.5$ is higher in non-equilibrium models than that for equilibrium models.
However in the equilibrium case, \fHeII strongly depends on density since \HeII
neutral fraction is directly set by number of recombinations and $f_{\rm HeII}
\propto \Delta^{0.6}$ (see Eq. \ref{eq:fHeII-Delta-relation}).  The \fHeII in
EQ-KS19-1.4 and NE-KS19-1.4 are very similar at $z=3.8$ (see Fig
\ref{fig:ionization-fraction-combine}).  However $T_0$ is higher and $\gamma$
is smaller for the NE-KS19-1.4 model than the EQ-KS19-1.4 model.  This can be
attributed to the fact that the amount of energy injected over a small redshift
range $\Delta z$ in the non-equilibrium case (NE-KS19-1.4) is more than that in
the equilibrium case (EQ-KS19-1.4) as \fHeII is higher in the former case.
The behaviour of TDR for equilibrium and non-equilibrium models 
seen in our case are qualitatively similar to those from \citetalias{puchwein2015} 
(their Fig. 2)\footnote{TDR in
\citetalias{puchwein2015} is plotted for \citetalias{haardt2012} UVB}.
However,Our method of calculating $T_0$ and $\gamma$ is different from that used in \citetalias{puchwein2015}.
The $T_0$ in \citetalias{puchwein2015} corresponds to the median temperature of all gas particles with densities within 5 percent of $\Delta=1$.
Whereas $\gamma$ in \citetalias{puchwein2015} is calculated by fitting a line between TDR at 
$(\Delta_1 = 10^{-0.5},T_{\rm median}(\Delta_1))$ and $(\Delta_2 = 1,T_{\rm median}(\Delta_2))$
where $T_{\rm median}(\Delta)$ is median temperature at $\Delta$.

\subsubsection{Evolution of $T_0$ and $\gamma$}
\label{subsec:T0-gamma-evolution}
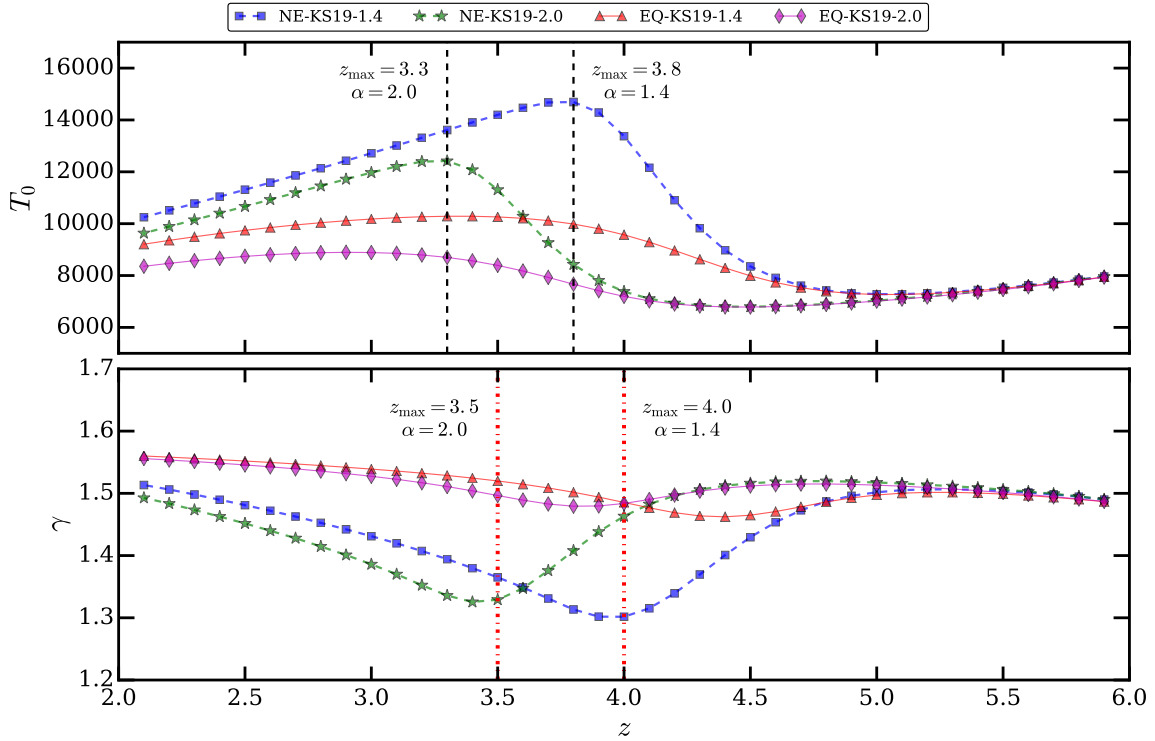
\InputFigCombine{Thermal_history_Two_SEDs.pdf}{150}{We show the evolution of $T_0$ 
    and $\gamma$ from \citetalias{khaire2018a} UVB compiled using two quasar spectral indices $\alpha = 1.4$ and $\alpha=2.0$.
    For each UVB, $T_0$ and $\gamma$ evolution is shown for equilibrium and non-equilibrium ionization evolution.
    Equilibrium model produces lower $T_0$ as compared to that from non-equilibrium case at $z < 5$. 
    This is because \fHeII is under predicted in equilibrium case as discussed in \S \ref{subsec:T0-gamma-evolution}. 
    (also see Fig. \ref{fig:ionization-fraction-combine}).
    On the other hand $\gamma$ is smaller for non-equilibrium models as compared to equilibrium models.
    This is due to the density independent photo-heating of the gas in non-equilibrium case. 
    The vertical dashed lines in top panel and the vertical dashed-dotted lines bottom panel 
    correspond to the redshift at which $d\log f_{\rm HeII} /dz$ and
    $d f_{\rm HeII}/dz$ are maximum respectively (see Fig. \ref{fig:fHeII-derivative})
    In all \citecode runs, we have assumed the initial $T_0$ ($= 7920$ K) and $\gamma$ ($= 1.55$) to be same at $z=6$.
}{\label{fig:T0-gamma-evolution-KS19}}

\InputFigCombine{Fraction_Derivative_Evolution_Combine.pdf}{140}{%
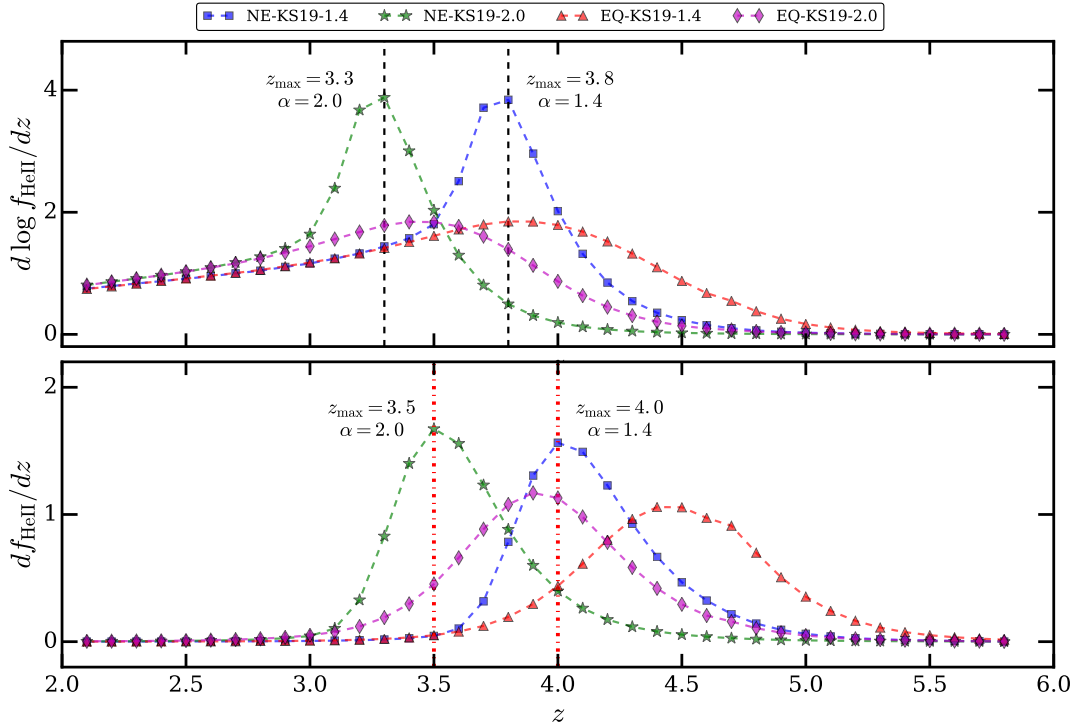
    Top panel shows the evolution of derivative of $\log f_{\rm HeII}$ with respect to $z$ (see Fig. \ref{fig:ionization-fraction-combine}).
    The peak in $d \log f_{\rm HeII}/dz$ correlates with peak in $T_0$ (dashed vertical lines) for all the UVB models (see Fig. \ref{fig:T0-gamma-evolution-KS19}).
    Bottom panel shows the evolution of derivative of $f_{\rm HeII}$ with respect to $z$.
    The peak in $d f_{\rm HeII}/dz$ correlates with epoch of minimum $\gamma$ (dash-dotted vertical lines) for all the UVB models.
    Thus \fHeII is a main driving factor affecting the $T_0$ and $\gamma$ evolution (see Fig. \ref{fig:T0-gamma-evolution-KS19}).
}{\label{fig:fHeII-derivative}}

In this section we discuss the evolution of $T_0$ and $\gamma$ for different UVB models discussed in \S\ref{subsec:uvb-models}.
Fig. \ref{fig:T0-gamma-evolution-KS19} shows the evolution of $T_0$ (upper panel) and
$\gamma$ (lower panel) for \citetalias{khaire2018a} UVB obtained using quasar SEDs with different
values of $\alpha$ (i.e $\alpha = 1.4$ and $2.0$).
For each UVB model, we show the $T_0$ and $\gamma$ evolution for equilibrium and non-equilibrium cases.
We now focus on the results for $\alpha=1.4$  i.e., models EQ-KS19-1.4 and NE-KS19-1.4 (the results are qualitatively similar for $\alpha=2.0$ models as well).
As expected based on \citetalias{puchwein2015} find that $T_0$ in the non-equilibrium model is higher as compared to that from the equilibrium model at any given epoch for $z < 4.5$.
As we discussed earlier, \fHeII is larger in the non-equilibrium model as compared to that in equilibrium at $3.5 < z < 5$.
As a result photo-heating (and hence temperature), which is proportional to \fHeII (see Eq. \ref{eq:heating-cooling-rate}), is more in the non-equilibrium model.
The maximum obtained temperature for the non-equilibrium model ($\sim 14500$ K) is substantially higher as compared to that of
equilibrium model ($\sim 10000$ K) at $3.5 < z < 5$.
Even though \fHeII for the non-equilibrium model is smaller than that for equilibrium model at 
$z < 3.5$ (see Fig. \ref{fig:ionization-fraction-combine}), it takes a long time (time scale $\sim$ Hubble time) for the gas to cool and reach equilibrium.
We also see more flattening (smaller $\gamma$) of the TDR in non-equilibrium model as compared to equilibrium model.
This is because the \fHeII (and hence photo-heating Eq. \ref{eq:heating-cooling-rate}) in the non-equilibrium model is independent of density (see Appendix \ref{app:semi-numerical}).

We next use the same figure to compare the $T_0$ and $\gamma$ evolution in the non-equilibrium models for the two values of $\alpha$ (1.4 and 2.0)
i.e., models NE-KS19-1.4 and NE-KS19-2.0.
The maximum $T_0$ obtained in NE-KS19-1.4 ($T_0 \sim 14500$ K at $z=3.7$) is 
higher by 17 percent than that from NE-KS19-2.0 ($T_0 \sim 12000$ K at $z=3.3$).
This is expected as a  higher value of $\alpha$ corresponds to less kinetic energy per 
photo-electron at $\lambda < 228 {\rm \AA}$. This excess energy 
would be less for the case of $\alpha = 2.0$ as compared to $\alpha = 1.4$.
The trend in $T_0$ vs $z$ seen in this figure can explain the kink we noticed 
in Fig.~\ref{fig:ionization-fraction-combine} for $\alpha = 1.4$ model.

The peak in $T_0$ evolution for non-equilibrium model occurs at lower redshift for higher $\alpha$ ($z=3.3$ for $\alpha=2.0$ and $z=3.7$ for $\alpha=1.4$).
This is consistent with the redshift evolution of \fHeII where we show that the redshift of reionization
is lower for higher value of $\alpha$ (see \S \ref{subsec:fHI-fHeII-evolution}).
It is interesting to note that for the non-equilibrium models, peak in $T_0$ evolution occurs at redshift 
where $f_{\rm HeII} \sim 0.055$ whereas the valley in $\gamma$ occurs at 
redshift where $f_{\rm HeII} \sim 0.25$ irrespective of UVB\footnote{This is even true 
for \citetalias{haardt2012} and \citetalias{khaire2018a} UVB with $\alpha=1.6$ and $1.8$}.
This also implies that the peak in the $T_0$ evolution and valley in the $\gamma$ evolution do not necessarily occur at the same redshift.
This behaviour of $T_0$ and $\gamma$ is correlated with the time-derivative of \fHeII.
We show the derivative of $\log f_{\rm HeII}$ (top panel) and \fHeII (bottom panel)
with respect to $z$ in Fig. \ref{fig:fHeII-derivative}.
One can see that the peak in $d \log f_{\rm HeII}/dz$ is nicely correlated with the peak in $T_0$ whereas
the epoch of the minimum $\gamma$ is correlated with $d f_{\rm HeII}/dz$.
The above trend is also seen in the case of equilibrium models. However peak in $T_0$ and $\gamma$ for equilibrium models
are quite broad.

\subsection{Evolution of the effective optical depth}
\label{subsec:tau-eff-evolution}
\InputFigCombine{HI_Tau_eff_Evolution.pdf}{175}{%
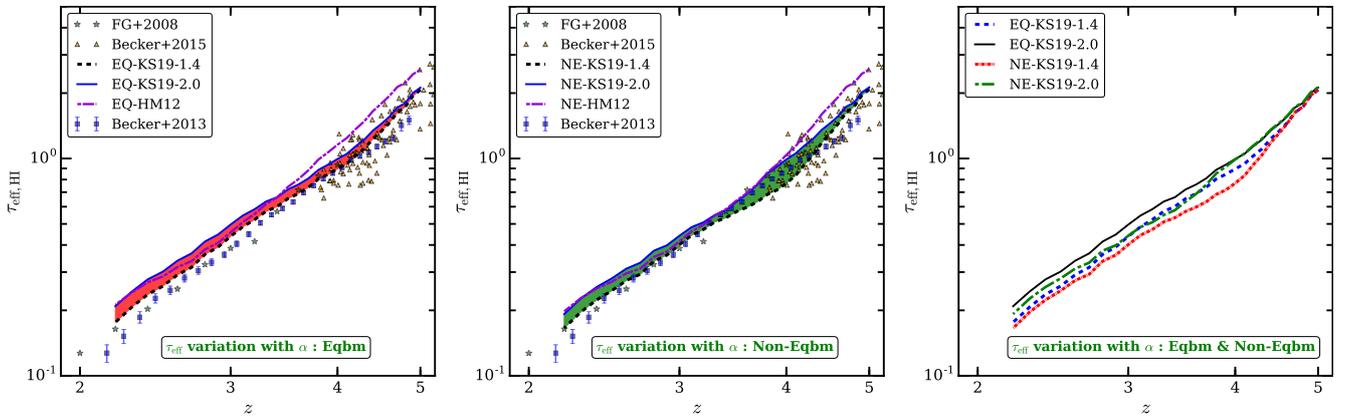
\GTW constraints from equilibrium (left panel) and non equilibrium models 
(right panel) using \taueffHI along with the measurements from \citet{becker2013}.
The observed \taueffHI from 
\citet[][black stars]{faucher2008}, \citet[][blue squares with errorbars]{becker2013} and
\citet[][orange triangles]{becker2015a} are shown in left and middle panel.
Left panel shows the \taueffHI evolution from equilibrium models for $\alpha = 1.4$ (black solid line)
and $\alpha = 2.0$ (blue solid lines) in \citetalias{khaire2018a} UVB model.
The red shaded region shows the uncertainty in \taueffHI due to uncertainty in $\alpha$ for equilibrium models.
\taueffHI evolution for $1.4 < \alpha < 2.0$ lies within red shaded region.
Middle panel is similar to left panel except that the \taueffHI evolution is shown for non-equilibrium models.
The green shaded region shows the uncertainty in \taueffHI due to uncertainty in $\alpha$ ($1.4 \leq \alpha \leq 2.0$) for non-equilibrium models.
The \taueffHI evolution matches well for the \citetalias{khaire2018a} non-equilibrium UVB model than that for 
\citetalias{khaire2018a} equilibrium UVB model.
In both panels, we also show the \taueffHI evolution for equilibrium and non-equilibrium model using \citetalias{haardt2012} UVB.
Right panel shows the comparison of \taueffHI evolution for equilibrium and non-equilibrium models.
The kink in the \fHI distribution in Fig. \ref{fig:ionization-fraction-combine} 
for NE-KS19-1.4 (red dotted line) manifest itself as a broad dip in the \taueffHI at $z\sim3.8$.
}{\label{fig:tau-eff-HI-evolution}}

\InputFigCombine{Median_HeII_Tau_eff_Evolution.pdf}{175}{%
    Each panel shows the evolution of \textit{median} \HeII effective optical depth (\taueffHeII) 
    from observations \citep[][red circles with errorbar]{worseck2018} and from simulations (blue stars with errorbars).
    The gray shaded regions correspond to 16$^{\rm th}$ and 84$^{\rm th}$ 
    percentile on simulated median \taueffHeII \citep[consistent with the corresponding range obtained by][]{worseck2018}.
    The median \taueffHeII and associated errorbars in simulation are calculated in a way 
    similar to that of \citet{worseck2018}.
    The \taueffHeII from simulation are shifted along $x$ axis by $\Delta z=0.01$ for clarity.
    The simulated \HeII \lya forest mimic the observations by accounting for the effects of 
    exposure time (varying SNR), HST-COS LSF, redshift path length and Poisson noise properties 
    that vary along different quasar sightlines \citep[Table 1 in][]{worseck2018}.
    Top (panel a, b, c and d) and bottom (panel e, f, g and h) rows show the median \taueffHeII
    evolution for equilibrium and non-equilibrium models (model name is given in each panel) respectively.
    The median \taueffHeII evolution from model EQ-KS19-1.8 (panel c) and NE-KS19-2.0 (panel h) are in good agreement with
    that from observations.
    Interestingly, our models produce the lower limits of \taueffHeII at $z>3$ 
    reasonably well whether we consider equilibrium  or non-equilibrium evolution.
    \PG{The large \taueffHeII scatter in our model at $z>3$ is due to fluctuations in the density field (UVB is assumed to be uniform).}
    The median \taueffHeII evolution at $z<2.8$ is sensitive to $\alpha$.
    The observed \taueffHeII evolution from \citet{worseck2018} is well reproduced by with $\alpha = 1.8$ for equilibrium models and
    $\alpha=2.0$ for non-equilibrium models.
}{\label{fig:tau-eff-HeII-evolution}}

\InputFigCombine{{tau_eff_CDF_Non_Eqbm_SED-2.0}.pdf}{175}{%
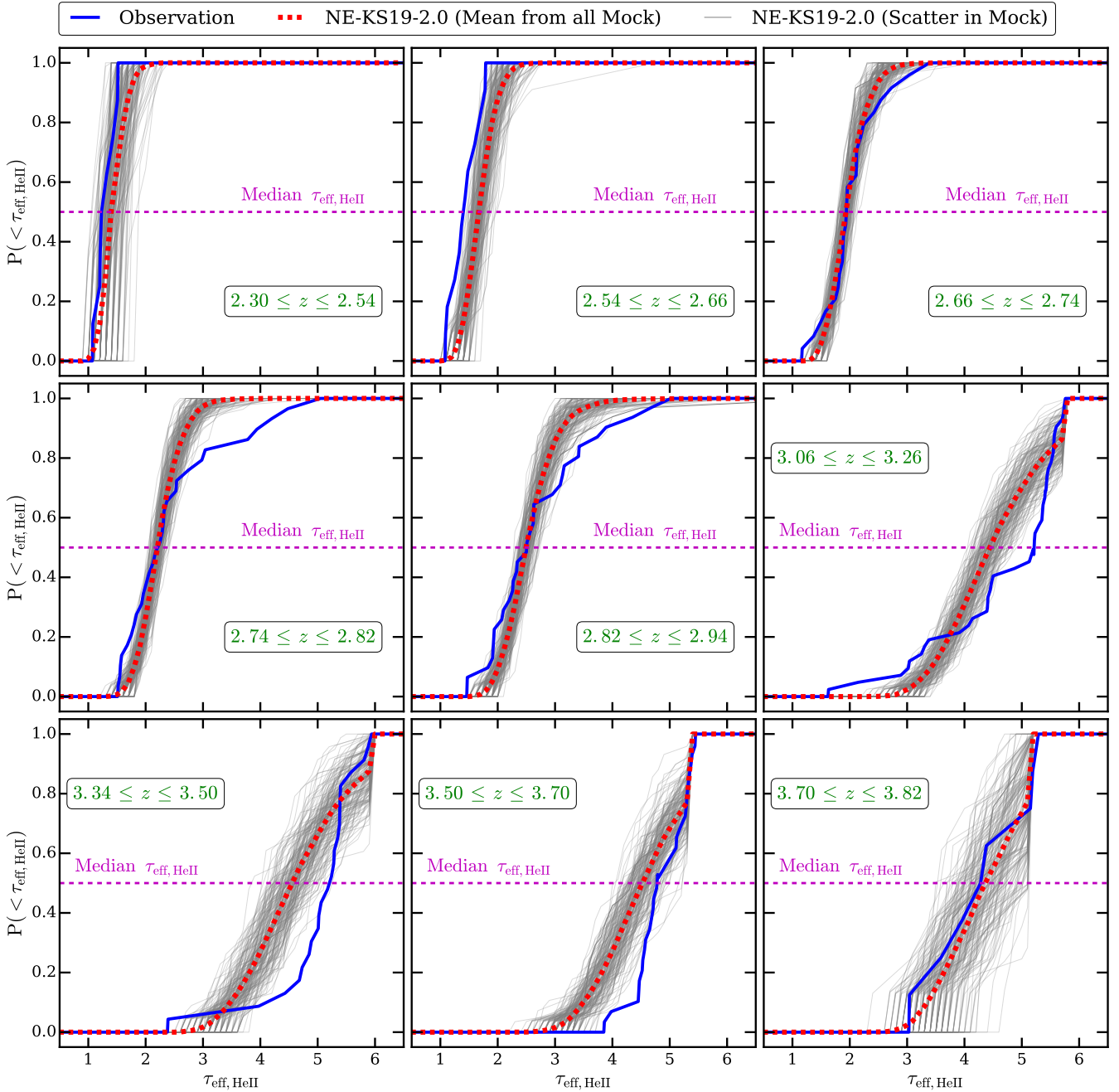
\PG{Each panel shows the comparison of cumulative distribution function of
\taueffHeII from observations (blue curve) with that from model NE-KS19-2.0
(uniform UVB model).  The model \taueffHeII CDF is obtained from 180 mock
samples and these are shown by thin gray curves. Each mock sample is mimicked to match
observations in terms of SNR, LSF, number of sightlines and redshift path
length. The mean \taueffHeII CDF from 180 mock samples these are shown by red dashed
curve in each panel.  The scatter in \taueffHeII is relatively well produced by
uniform UVB model at $z < 2.74$.  However, in high-$z$ bins our models have more 
low optical depth regions compared to the observations. This lack of good matching between 
the two CDF may favour additional mechanisms like fluctuations in UVB and/or temperature
in addition to what we get from density fluctuations only.}
}{\label{fig:tau-eff-HeII-cdf-evolution}}
In this section, we present the comparison of our model predictions with the observations of \lya forest.
To generate the forest, we shoot random sight lines through our simulation box and calculate the overdensity
($\Delta$), peculiar velocity ($v$) and temperature ($T$) along these skewers
\citep{trc2001,hamsa2015,gaikwad2017a}.
Since we calculate the temperature of each particle in the post-processing of \gtwo
using \citecode, we need to correct for the pressure smoothing effects.
We follow the procedure given in \citetalias{gaikwad2018} to generate the \lya 
forest spectra along these sightlines.
In \citetalias{gaikwad2018}, we have shown that our method is accurate within 20 per cent
with that from self-consistent \gthree simulation\footnote{The effective optical depth 
from \gtwo + \citecode matches within $1.75$ percent with that from \gthree.}.
We also account for the line spread function (Gaussian with FWHM $\sim 6$ \kmps),
finite SNR ($\sim 20$) for generating the mock \lya forest spectra.
In the case of \HeII \lya forest we have used the line spread function, signal to noise
ratio and pixel sampling as appropriate for the HST-COS spectra.

The effective optical depth, $\tau_{\rm eff} = - \log \langle e^{-\tau} \rangle$, where the angle 
brackets indicate averaging, is a robust quantity that can be easily derived from 
observations and simulations for both \HI and \HeII.
Fig. \ref{fig:tau-eff-HI-evolution} shows the evolution of \HI effective optical depth (\taueffHI)
 with redshift for different UVB models.
The observed \taueffHI data points from \citet{faucher2008,becker2013,becker2015a} are also
shown in the figure. 
To calculate \taueffHI from simulations, we use boxes at $z=2.1$ to $5.0$ with $\Delta z= 0.1$.
For each UVB model, we calculate \taueffHI using 500 sight lines at each redshift.

The left panel of Fig. \ref{fig:tau-eff-HI-evolution} shows the results for the equilibrium  models for \citetalias{khaire2018a} UVB
and \citetalias{haardt2012}.
We can see that the \taueffHI is consistently larger for $\alpha=2.0$ as compared to that for $\alpha=1.4$.
This is expected as $\tau_{\rm eff} \propto T_0^{-0.7}$ because of the temperature dependence of the recombination rate coefficient and $T_0$ is consistently 
larger for smaller $\alpha$ (as we have already seen in Fig. \ref{fig:T0-gamma-evolution-KS19}).
We show the \taueffHI for $1.4 < \alpha < 2.0$ as the red shaded region with the upper and lower envelops of this
shaded region correspond to $\alpha = 2.0$ and 1.4 cases respectively.
Interestingly, \taueffHI for the \citetalias{khaire2018a} and \citetalias{haardt2012} equilibrium models are consistently higher 
than those from observations at $z < 4$, with the mismatch being maximum at the lowest redshifts.
Thus for our simulations, to match the observed \taueffHI, we would require \GHI which is 18 percent lower than values obtained for \citetalias{khaire2018a}.
    
The middle panel in Fig. \ref{fig:tau-eff-HI-evolution} is similar to the left one except that the \taueffHI evolution is shown for 
 non-equilibrium UVB models for \citetalias{khaire2018a} and \citetalias{haardt2012}.
Since $T_0$ is consistently higher for the non-equilibrium models as compared to the equilibrium models,
\taueffHI for the non-equilibrium model is lower.
Interestingly, this makes the \taueffHI evolution for non-equilibrium case match better with the observations in the range 
from $z=5$ to $z=2.5$.
This also implies that, for a given set of cosmological parameters, the observed 
\taueffHI would lead to different \GHI depending on whether one uses the 
equilibrium or the non-equilibrium model while comparing with the data. 
The kink we noticed in the predicted \fHI distribution in Fig. \ref{fig:ionization-fraction-combine} 
for NE-KS19-1.4 manifest itself as a broad dip in the \taueffHI curve as evident in the
lower envelop at $z\sim3.8$ (right panel in Fig. \ref{fig:tau-eff-HI-evolution}).
Similar broad dip in \taueffHI is also seen in \citetalias{puchwein2015} for \citetalias{haardt2012} UVB.

Moving on to \HeII, we show the evolution of \taueffHeII with redshift for different UVB models  for equilibrium and non-equilibrium considerations
in Fig.~\ref{fig:tau-eff-HeII-evolution}. For comparison, we use the observed median, 16$^{th}$ and 84$^{th}$ percentile of \taueffHeII over different redshift bins listed in table 4 of \citet{worseck2018}. These were obtained with quasar
spectra observed with different  spectral resolution (i.e line spread functions, LSFs) and exposure times (i.e varied SNR). In order to make realistic comparisons, while computing the distribution of \taueffHeII from our simulations  we used appropriate combinations of LSF and SNR. Inferred flux or optical depth
at the bottom of the saturated lines  will be dominated by photon noise originating from the background subtraction. We also model this effect carefully.

The shaded regions in Fig~\ref{fig:tau-eff-HeII-evolution} are our model predicted regions covering 16$^{th}$ and 84$^{th}$ percentile of \taueffHeII. When we consider the equilibrium models (i.e upper panels in Fig.~\ref{fig:tau-eff-HeII-evolution}), the UVB computed using $\alpha = 1.8$ (panel c) reproduces the observed median and the scatter very well for $z<2.8$. This model also captures the flattening of \taueffHeII with large scatter seen in observations at high-$z$ very well.  From the figure 3 of \citet{worseck2018} we can see that the equilibrium simulations using uniform UVB of \citet{puchwein2018} produce much sharper evolution of \taueffHeII as a function of $z$ compare to our models. This is probably related to
sharper increase in \GHeII produced by the UVB of \citet{puchwein2018} compared to those used in our models. It is also clear that the scatter we find in \taueffHeII is much more than what has been found for constant \GHeII models considered by \citet{worseck2018}. Thus our models seem to produce a reasonable agreement to the data unlike the uniform background models considered in \citet{worseck2018}. This is not surprising as constraints from He~{\sc ii} measurements were used while constructing the UVB used in our study \citep[see][]{khaire2017}. From Fig~\ref{fig:tau-eff-HeII-evolution} we notice that at $z>3$, 16$^{th}$ percentile \taueffHeII from our models may be lower than what is observed (even when we match the limiting values in two redshift bins observational limits are higher in the middle two $z$ bins).

Next we consider the non-equilibrium models (lower panels in Fig.~\ref{fig:tau-eff-HeII-evolution}). As expected \taueffHeII obtained at $z<2.8$ are lower than that obtained for the corresponding equilibrium cases. Thus in order to match the observed data we require UVB obtained with slightly larger value of $\alpha$ (i.e $\alpha = 2$, panel h). On the other hand non-equilibrium models provide slightly higher values of \taueffHeII at $z>3$ compared to the corresponding equilibrium models. However even in this case observed limits in two redshift bins are higher than that predicted by our models.

In summary, both equilibrium and non-equilibrium models reproduce the observed
trend of \taueffHeII as a function of $z$.  
\PG{As we use uniform UVB, the large \taueffHeII scatter produced by our models at $z>3$ is mainly due to fluctuations in the 
density field and evolution in \GHeII}\footnote{The contribution of 
temperature fluctuations is sub-dominant.} (also see Appendix \ref{app:tau-eff-HeII-scatter}). 
For the constant \GHeII value used by \citet{worseck2018}, we expect \taueffHeII scatter to be smaller than what is produced by our models. The UVB used by \citet{puchwein2018}
produces larger \taueffHeII than what is observed while producing large scatter. 
It is important to note that the 
effect of simulation box size, mass resolution \citep{bolton2017}, finite noise, spectral resolution and
redshift path length in the observed spectra have weaker effect on \taueffHeII scatter at $z>3$\footnote{We
have tested the effect of box size and mass resolution on \taueffHeII in Sherwood simulation suite 
\citep{bolton2017}.}.
Our study shows the observed range
of \taueffHeII at $z<2.8$ can be used to place constraints on the quasar
spectral index $\alpha$ \citep[also see][]{khaire2017}. The non-equilibrium
models require higher value of $\alpha$ compared to the equilibrium models. Our
models produce the enhanced scatter in \taueffHeII at $z>3$ with
non-equilibrium models producing slightly higher values. However, if the
observed \taueffHeII at $z>3$ are not dominated by small number statistics then
our model predicted 16$^{th}$ quartile values (both equilibrium and
non-equilibrium case) may be smaller than what has been observed. 
We show the comparison of mean \taueffHeII evolution from our models with that from 
observations \citep{heap2000,zheng2004,fechner2006,syphers2014,worseck2016} in
Appendix \ref{app:mean-tau-eff-HeII-evolution} (see Fig. \ref{fig:mean-tau-eff-HeII-evolution}).
Qualitatively our mean \taueffHeII evolution is consistent with that from 
\citet{puchwein2015,puchwein2018}.

\PG{Fig. \ref{fig:tau-eff-HeII-cdf-evolution} shows the comparison of
    cumulative distribution function (CDF) of \taueffHeII from observations with that
    from model NE-KS19-2.0 (uniform UVB model) at $2.30 \leq z \leq 3.82$.  We
    show \taueffHeII CDF for NE-KS19-2.0 model since the median \taueffHeII
    evolution from this model is in good agreement with observations (Fig.
    \ref{fig:tau-eff-HeII-evolution}). The model \taueffHeII CDF in Fig.
    \ref{fig:tau-eff-HeII-cdf-evolution} is obtained from 180 mock samples.
    Each mock sample is mimicked to match observations in terms of SNR, LSF,
    number of sightlines and redshift path length.  The scatter in \taueffHeII
    is relatively well produced by uniform UVB model at $z < 2.74$.  However,
    the \taueffHeII scatter from uniform UVB model is smaller than the
    observed CDF for higher redshift bins. Thus, while we could produce larger scatter
    in median \taueffHeII covering the observed scatter using an uniform UVB, the analysis
    presented here suggests that we may need additional sources of scatter like 
    fluctuations in the UVB and/or temperature to explain the observed CDF. As mentioned
    before consideration of non-equilibrium evolution does not produce significantly large
    scatter than the equilibrium models.}
%


\vspace{5mm}
As discussed above our models require small fine-tuning of the ionizing background (mainly the normalization) to match the \taueffHI as a function of $z$. 
For example, one may have to increase the value of \GHI (by $\sim 18$ percent)
which may result in the reduction of \GHeII through radiative transport effects
as shown in \citet{khaire2013}. This will in turn increase the \taueffHeII. We
plan to carry out such a self-consistent modelling in near future. Note
\citet{worseck2018} explained the observed \taueffHeII as a function of $z$
using fluctuating UVB models. Moreover, the fluctuations in the temperature due
to residual heating from \HeII reionization could be important in the redshift of
interest \citep{keating2018}. Given that the simulation box size used in our
study is small, modelling the effect of fluctuating UVB and temperature will 
not be possible in the present case.
However, our study emphasizes the non-equilibrium effects
that will be important even when one uses fluctuating UVB and temperature.
We would like to emphasize here that even for models that includes 
fluctuating UVB, our post-processing code will be handy in incorporating non-equilibrium
effects.

\subsection{\GHI constraints}
\label{subsec:Gamma-HI-constraints}
\InputFigCombine{Gamma_HI_Constraints.pdf}{160}{%
Each panel shows constraint on \GTW from this work.
The \GTW constraints from \citet{becker2013} are shown for $\gamma = 1.2, 1.4$ and $1.6$ 
by green, blue and orange solid lines respectively.
Left panel shows \GTW constraints for $\alpha = 1.4$ (red stars) and $\alpha = 2.0$ 
(magenta squares) \citetalias{khaire2018a} equilibrium UVB model.
The red shaded region shows the uncertainty in \GTW due to uncertainty in $\alpha$ ($1.4 \leq \alpha \leq 2.0$).
Right panel is similar to left panel except the \GTW constraints are shown for non-equilibrium model.
The green shaded region encompasses allowed \GTW for $\alpha$ in the range $1.4 \leq \alpha \leq 2.0$.
The redshift evolution of \GTW constraints is different in equilibrium and non-equilibrium models.
}{\label{fig:Gamma-HI-constraints}}

One or more of observed \taueffHI, flux probability distribution function or flux power spectrum  of \lya absorption are used to constrain
\GHI \citep{bolton2007,faucher2008c,becker2013,gaikwad2017a}. However for a given density field the above mentioned quantities also depend on the gas temperature
and thermal history through pressure smoothing \citep{gnedin1998,kulkarni2015}. Thus it is a usual procedure to use simulations with 
varied thermal histories to measure \GHI as a function of $z$ \citep{becker2011,becker2013}.
Here we ask how the inferred \GHI
for a given observed
\taueffHI depends on whether one uses equilibrium or non-equilibrium evolutions.

    Typical steps followed in constraining \GHI involves, 
    \begin{enumerate}
        \item generating \lya forest in simulation for a given UVB (and hence ionization and thermal history) using either equilibrium or non-equilibrium conditions, 
        \item compare the observed \taueffHI with predictions from the simulation, 
        \item rescaling the simulated \GHI at a given epoch in the post-processing step to make the 
              \taueffHI from simulations match with that from observations and 
        \item repeating the step (i) to (iii) for different UVB.
    \end{enumerate}
    Note that we have implicitly assumed here that changing the value of \GHI locally does not affect
    the thermal history at redshifts of our interest. Also at the redshift of our interest collisional 
    ionization is sub-dominant.
    This is a reasonable approximation if the applied
    scaling is at the level of few percent.
    The first step in obtaining different thermal histories is computationally expensive 
    as explained in \S\ref{sec:simulation}. In our case we use \citecode to generate the
    simulations for an assumed UVB.

We use \citet[][their table 2]{becker2013}  \taueffHI measurements 
to constrain \GHI in 7 redshift bins centred at $z=2.4-4.8$ with $\Delta z = 0.4$.
\citet{becker2013} have tabulated the
\taueffHI measurements from observations accounting for the possible
systematics such as metal contaminations, continuum placement uncertainty,
correction for the presence of optically thick absorbers etc \citep[also see][]{faucher2008}. 
For each of our models we constrain \GHI by finding the root (\GHI value) that matches model \taueffHI with observed \taueffHI at a
given redshift.  We use \textit{scipy's} optimize module to find the root
(function tolerance set to $10^{-9}$).  We use 2000 sight lines to calculate the
\taueffHI from simulation at any given redshift.  We splice sight lines to match
the observed redshift path length ($\Delta z = 0.4$).  
We add a Gaussian random noise of SNR$=20$
to the simulated \lya forest spectra.

The left and right panels in Fig. \ref{fig:Gamma-HI-constraints} show \GTW constraints 
(\GHI in $10^{-12} \: {\rm s}^{-1}$) for EQ-KS19 and NE-KS19 models respectively from this work.
In each panel, we show the \logGTW for two values of $\alpha = 1.4$ and $2.0$.
First, we consider the variation in \logGTW due to variation in $\alpha$ for a given model
i.e., either EQ-KS19 or NE-KS19.
The best-fit \logGTW values are systematically higher for smaller values of $\alpha$.
This is consistent with the evolution of $T_0$ in Fig. \ref{fig:T0-gamma-evolution-KS19} where $T_0$ 
is consistently larger for smaller values of $\alpha$ irrespective of whether we use EQ-KS19 or NE-KS19 model.
Since for a given \taueffHI we have $\Gamma_{\rm HI} \propto T_0^{-0.7}$,
the constrained \logGTW values are systematically higher for larger $\alpha$.
The same explanation also holds when we compare the \GTW constraints from models with same $\alpha$ 
but different ionization evolution i.e., EQ-KS19-1.4 (EQ-KS19-2.0) and NE-KS19-1.4 (NE-KS19-2.0) models. 
Since $T_0$ is systematically larger for NE-KS19-1.4 model than in the EQ-KS19-1.4 model,
\logGTW in NE-KS19-1.4 is consistently smaller in EQ-KS19-1.4 model.
We also find that the \GHI constraints are relatively insensitive to the choice of initial $T_0$, $\gamma$ we use at 
$z=6$ in \citecode (see Appendix \ref{app:uncertainty-in-T0-gamma-evolution} for details).

We now consider the allowed range in \logGTW as shown by red and green shaded regions
in the left and right panels of Fig. \ref{fig:Gamma-HI-constraints} respectively.
We also show \logGTW measurement from \citet{becker2013} for $\gamma=1.2,1.4$ and $1.6$  
in each panel of Fig. \ref{fig:Gamma-HI-constraints}.
The derived \logGTW range and its redshift dependences for NE-KS19 model (green shaded region in right panel) at 
$3.5 < z < 5$ are different compared to those obtained for EQ-KS19 (red shaded region in left panel) models.
This is a direct consequence of difference in $T_0$ and $\gamma$ evolution between
these models.
Fig. \ref{fig:Gamma-HI-constraints} also shows that the minimum in \logGTW at 
$z \sim 3.5$ is more prominent in non-equilibrium models than that in equilibrium models. 

The difference in the derived shape of $\Gamma_{12}$, in particular a prominent 
valley observed at $z\sim 3.5$ can naively imply that there are two kinds of 
sources, one dominating the UV background at $z<3.5$ and another at $z>3.5$.
For example, it will not be straightforward to obtain such a valley-like feature 
in the UVB models using only one population of sources such as quasars 
\citep[see for e.g][]{madau2015, khaire2016}. One will be forced to invoke 
stronger evolution in $f_{\rm esc}$ than the one assumed in these models.

To summarize, the widely used technique to obtain \GHI from observed \taueffHI 
distribution calibrated by simulations can result in \GHI evolution that is very 
different from the intrinsic distribution depending upon whether one uses
equilibrium and non-equilibrium models. In particular in the redshift range
where He~{\sc ii} reionization takes place it will be important to
simultaneously model \taueffHI, \taueffHeII and $T_0$ and $\gamma$ evolutions
to place constraints on the nature of UV ionizing background. 
\citet{becker2013} constrained $\Gamma_{\rm HI}(z)$ using a range of $T_0-\gamma$ 
combinations whose redshift evolution need not be realized in a self-consistent 
model. Our post-processing code \citecode allows us to consider self-consistent 
$T_0-\gamma$ for assumed \HeII reionization model under equilibrium and non-equilibrium
conditions while constraining \GHI.
It is clear from the discussions
in the previous section that in a physically motivated model the redshift
evolution of $T_0$ and $\gamma$ can allow us to distinguish between equilibrium
and non-equilibrium cases.  However, both these quantities are not directly
measured but derived from the observed properties of \lya forest in the
framework of simulations. In the following section we try to see whether one
will be able to recover the intrinsic evolution of $T_0$ and $\gamma$ using
equilibrium models as done in the literature.

\subsection{Recovery of $T_0$ and $\gamma$ in non-equilibrium recovery from equilibrium model}
\label{subsec:T0-gamma-recovery}

\InputFigCombine{T0_gamma_constraint.pdf}{180}{%
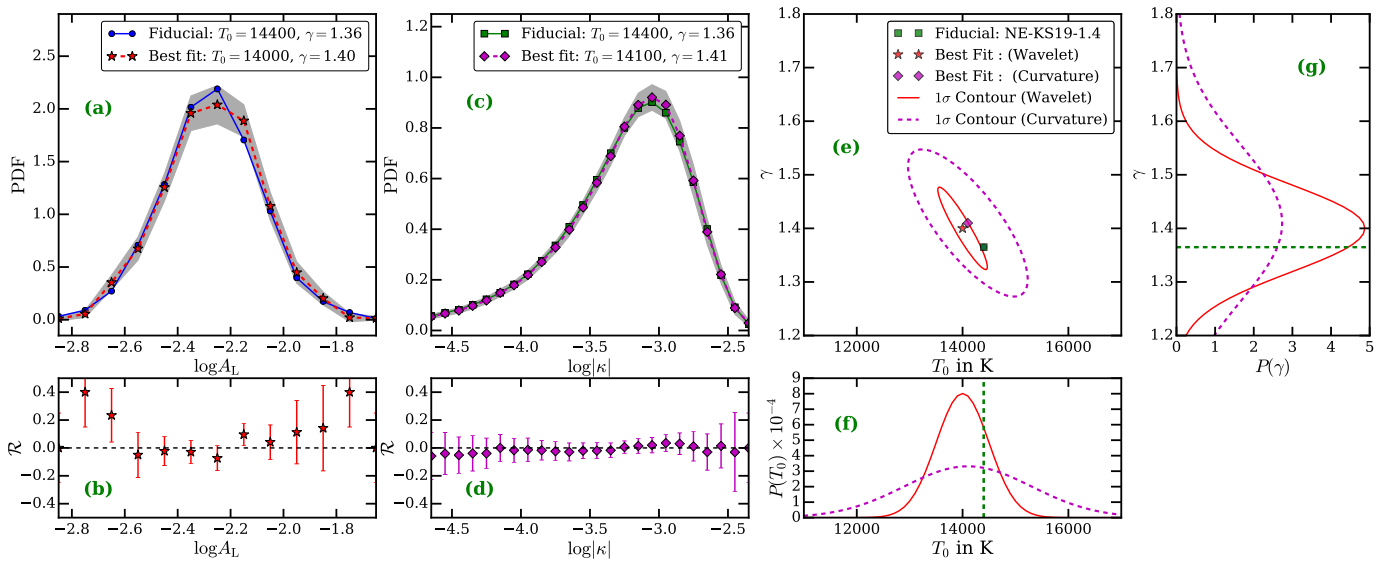
    Panel {\bf a} shows the wavelet PDF for fiducial model (NE-KS19-1.4, blue circles) 
    and best fit model (EQ-KS19-1.4, red stars).
    We vary the thermal history in EQ-KS19-1.4 model to obtain the different $T_0$ and $\gamma$ combination (see \S \ref{subsec:T0-gamma-recovery}).
    The best fit model corresponds to minimum $\chi^2$ (or maximum likelihood) between fiducial and model wavelet PDF.
    The corresponding residuals between fiducial and model wavelet PDF are shown in panel {\bf b}.
    Panel {\bf c} and {\bf d} are similar to panel {\bf a} and {\bf b} respectively except for PDF of curvature statistics.
    Joint $1\sigma$ contours (maximum likelihood) from wavelet (red solid) and curvature (magenta dashed) statistics are shown in panel {\bf e}.
    The marginalized distributions for $T_0$ and $\gamma$ are shown for each statistics in panel {\bf f} and {\bf h}.
    Both statistics recover the fiducial (NE-KS19-1.4, shown by green dashed line in panel {\bf f} and {\bf h}) 
    $T_0$ and $\gamma$ within $1\sigma$ uncertainty. 
   All plots are shown at $z=3.6$ and for KS19-1.4 UVB models.
}{\label{fig:T0-gamma-recovery}}

The thermal history of the IGM provides an indirect way of studying \HeII reionization.
In the literature, $T_0$ and $\gamma$ are constrained from observations
by comparing with predictions of simulations having a wide range of 
thermal histories. However, most of these simulations do not consider non-equilibrium ionization conditions.
As shown in \S \ref{subsec:T0-gamma-evolution}, the thermal history of IGM can 
be considerably different in equilibrium and non-equilibrium ionization models and the non-equilibrium effects could be quite important during \HeII reionization.
Hence it is important to investigate whether the thermal history in non-equilibrium models can be
recovered using an equilibrium model.

We assume the underlying ``true'' model to be given by NE-KS19-1.4. We concentrate on $z = 3.6$ which is well within period of the \HeII reionization. The thermal parameters at this redshift for the NE-KS19-1.4 are $T_0 = 14400$ K and $\gamma = 1.36$. 
Next, we generate the different thermal histories for the EQ-KS19-1.4 model using the \citecode.
We vary the photo-heating rates as $\epsilon_{\rm X} = a \: \Delta^{b} \: \epsilon_{\rm X,KS19}$ \citep{becker2011},
where $\epsilon_{\rm X,KS19}$ is photo-heating rate of species ${\rm X} \equiv [{\rm HI,HeI,HeII}]$
in the EQ-KS19-1.4 model \citep[also see][for different approach]{onorbe2017}. 
A variation in factor $a$ (or $b$) corresponds to change in $T_0$ ($\gamma$, respectively) with small variation
in $\gamma$ ($T_0$, respectively). 
We varied factor $a$ and $b$ such that the $T_0$ varies from 6000 K to 20000 K in steps of 2000 K and
$\gamma$ varies from 1.1 to 2.0 in steps of 0.1. In total we generate $8 \times 10 = 80$ different 
thermal histories from EQ-KS19-1.4 model at $z=3.6$. 
Note unlike the self-consistent models the pressure smoothing we apply for each particle will depend on its temperature
at $z \sim 3.6$ and may not capture the effect of thermal history. 
However as shown by \citetalias{gaikwad2018} the effect is of the order of $< 20$ percent in the two PDFs discussed here 
(see section 4.4 and 4.5 in \citetalias{gaikwad2018})

For each of these simulations (including the underlying ``true'' model), we shoot $1000$ random sight lines and compute \lya forest spectra.
We adjust \GHI in all the models such that the mean transmitted flux is same.
For fair comparison, we account for the effects of finite SNR and LSF in the \lya forest spectra for all the models.
In order to constrain $T_0,\gamma$ from \lya forest, we use two 
statistics namely wavelet statistics \citep{zaldarriaga2002,theuns2002b,lidz2010,garzilli2012,gaikwad2018} 
and curvature statistics \citep{becker2011,boera2014,hamsa2015,gaikwad2018}.
These statistics have the property that a large value of temperature corresponds to smaller
values of curvature and wavelet amplitudes. The method to calculate these statistics is identical to that
of \citetalias{gaikwad2018}. 
We compute $\chi^2$ (and maximum likelihood $\mathcal{L} \propto e^{-\chi^2/2}$)
between the assumed ``true'' non-equilibrium model and the equilibrium model 
(for each thermal history) for both the statistics. 
The best-fit model corresponds to a minimum value of $\chi^2$ (or a maximum value of $\mathcal{L}$). 
The statistical uncertainty in $T_0$ and $\gamma$ corresponds to $\chi^2 = \chi^2_{\rm min} + \Delta \chi^2$ 
where $\Delta \chi^2 = 2.30$ for $2$ ($T_0$ and $\gamma$ being free parameters) degrees of 
freedom \citep{avni1976,press1992}\footnote{It is assumed that the errors are distributed normally}.

Fig. \ref{fig:T0-gamma-recovery} shows  the recovery of the ``true'' (NE-KS19-1.4) $T_0$ and $\gamma$  using
EQ-KS19-1.4 models with different thermal histories.
Panels a and c show the PDFs of the wavelet and curvature statistics respectively for the ``true'' and best-fit models.
The corresponding residuals between the ``true'' and best-fit models are shown in panels b and d.
The reduced $\chi^2$ between the ``true'' and best-fit models for wavelet and curvature statistics are $1.61$ and $1.19$
respectively. 
Panel e shows the joint $1\sigma$ constrain on $T_0-\gamma$ from curvature 
(magenta dashed curve) and wavelet PDF (red solid curve).
The assumed ``true'' $T_0$ and $\gamma$ lie within $1\sigma$ contours for both the statistics.
The marginalized distributions  (panel f and h) too show that the equilibrium models can 
recover the non-equilibrium $T_0,\gamma$ within $1\sigma$ statistical uncertainty.
\PG{The $T_0-\gamma$ constraints from wavelet statistics are better than the curvature statistics because
we do not smooth the wavelet amplitude i.e., we use 
the Eq. 2 to Eq. 6 from \citet{lidz2010} but we do not use Eq. 7 of \citet{lidz2010}.}

In a realistic scenario, one would expect the $T_0$ and $\gamma$ evolution to be given by the
non-equilibrium models, and hence one should use non-equilibrium simulations to constrain the thermal parameters.
Unfortunately, the non-equilibrium models are computationally more expensive than the corresponding equilibrium models, hence exploring a large $T_0-\gamma$ parameter space is often impractical with non-equilibrium models.
The above result indicates that the underlying non-equilibrium thermal parameters can be recovered by varying the
thermal history in equilibrium models, \emph{as long as the \taueffHI in these models are the same}.
We also show the effect of matching \taueffHI on FPDF and FPS statistics of \HI \lya
forest in Appendix \ref{app:lya-statistics}. 
In the literature the $T_0$ and $\gamma$ are constrained by allowing the mean flux to vary \citep[see][]{becker2011}.
If this is true at all $z$ then the derived $T_0$ and $\gamma$ evolution will reflect whether the underlying evolution
is equilibrium or non-equilibrium case.

We caution the reader that our conclusions are based on our model where 
the large scale fluctuation in \GHeII, $T_0$ and $\gamma$  expected for fluctuating
UVB are not be modelled accurately as the box size (10 $h^{-1} {\rm \cmpc}$) used is small. Also subtle
variations in the pressure smoothing due to small changes in the thermal history are
not captured in CITE. 
A self-consistent radiative transfer simulation with large box size, sufficient 
mass resolution and non-equilibrium ionization solver is needed to capture 
these effects \citep{aubert2008,rosdahl2013,gnedin2014,kulkarni2018}.
However these simulations may not be flexible enough to probe large parameter spaces.

\section{Summary}
\label{sec:summary}
The cosmological hydrodynamical simulations used
to study the properties of \lya forest in the literature
implement the equilibrium ionization evolution for a given UVB model.
However, as we have shown here the assumption of photo-ionization equilibrium 
may not be valid at $2 \leq z \leq 4$ where \HeII reionization is in progress.
In addition, the quasar spectral shapes in extreme UV used in the modelling of UVB are observationally ill-constrained.
All this can lead to systematics in the derived quantities such as thermal state of the IGM quantified by $T_0$ and $\gamma$,
\HeII fraction (\fHeII), \HI fraction (\fHI) and \HI photo-ionization rate (\GHI) that are related to \HeII reionization process.
In this work, we implement the non-equilibrium ionization evolution in our post-processing tool ``\citefullform'' (\citecode).
Since \citecode is post-processing module that works on output of \gtwo (SPH) simulation, we can efficiently simulate the effect of wide range in UVB
for equilibrium and non-equilibrium models.
We show the consistency of equilibrium and non-equilibrium ionization 
evolution obtained using \citecode with those from self-consistent \gthree simulations \citep{puchwein2015,gaikwad2018}.
Having established this, we explore the effect of unknown quasar UV spectral index
on the derived properties of \HeII and \HI absorption as a function of redshift.
We summarize our findings as follows,

For a given ionization scenario (equilibrium or non-equilibrium), \fHeII 
is systematically smaller for UVB models obtained using flatter quasar SEDs at $2 \leq z \leq 6$.
This suggests that the redshift of \HeII reionization ($z_{\rm reion}$) 
strongly depends on quasar spectral indices used in the UVB models such that \HeII reionization is earlier 
($z_{\rm reion}$ is larger) when flatter quasar SEDs are used.
The extent of \HeII reionization ($\Delta z$) is smaller for non-equilibrium models
than that from equilibrium models. $\Delta z$ is relatively insensitive to variation
in quasar spectral indices.

The globally volume averaged \HeII fraction (\fHeII) in non-equilibrium model 
evolves rapidly from $1$ to $\sim 10^{-2}$ over a small redshift interval $\Delta z = 0.85$. 
Whereas the corresponding change in \fHeII for equilibrium model occurs over 
a larger redshift interval $\Delta z = 1.3$ (see Fig. \ref{fig:ionization-fraction-combine}). 

For a given UVB, $T_0$ in non-equilibrium model is consistently  larger than that in 
equilibrium model at $2 \leq z \leq 6$. 
Whereas in the same redshift range, $\gamma$ is consistently smaller 
in non-equilibrium model than that in equilibrium model \citep{puchwein2015}.
In all models, we notice that the epoch of minimum $\gamma$ occurs earlier than the epoch of maximum $T_0$.
The epochs of maximum $T_0$ and minimum $\gamma$ are well correlated with
derivatives of \fHeII with respect to redshift (see Fig. \ref{fig:fHeII-derivative}).
On the other hand, for a given ionization evolution, $T_0$ is consistently 
larger for models using UVB obtained with flatter quasar spectral index.
The epoch of maximum $T_0$ (and minimum $\gamma$) is earlier for simulations
using UVB compiled with flatter quasar SEDs (Fig. \ref{fig:T0-gamma-evolution-KS19}).
This evolution of $T_0$ and $\gamma$ for all models (equilibrium, non-equilibrium 
and different quasar spectral indices) is consistent with the evolution of \fHeII.

For a given set of cosmological parameters and given \GHI, we find that the 
\taueffHI at any $z$ is consistently smaller in non-equilibrium model as compared
to equilibrium model (Fig. \ref{fig:tau-eff-HI-evolution}). 
The predicted \taueffHI at any $z$ is systematically smaller for 
models using UVB generated with flatter quasar SED. 
This is because $T_0$ is larger in non-equilibrium and/or for UVB generated with 
flatter quasar SEDs.

We show the observed median \taueffHeII and 16$^{th}$ and 84$^{th}$ percentiles
regions are well reproduced by both equilibrium and non-equilibrium models.  In
particular we find that the observed \taueffHeII at $z<2.8$ can be used to
constrain $\alpha$ values. The equilibrium models require slightly smaller
values of $\alpha$ to reproduce the observed range.  \PG{While our models
    produce 16$^{th}$, 50$^{th}$ and $84^{th}$  percentile of \taueffHeII
    relatively well with observations, to understand the scatter in \taueffHeII
    we show comparison of observed CDF with model CDF. The scatter in
    \taueffHeII is relatively well produced by uniform UVB models at $z <
    2.74$. However, the observed \taueffHeII scatter at higher redshift is
    larger than those in uniform UVB models. Thus, this work suggests that we
    may need additional sources of scatter in \taueffHeII such as fluctuations
    in UVB and /or temperature to explain the observed \taueffHeII CDF.}
    
We estimate the \GHI in equilibrium and non-equilibrium models with UVB
obtained using quasar spectral indices $\alpha = 1.4$ and $2.0$ and by matching
the \taueffHI from our models with that from \citet{becker2013} observed data.
The \GHI estimates are systematically larger for UVB models with higher quasar
spectral index at $2.4 \leq z \leq 4.8$ due to dependence of $T_0$ evolution on
quasar spectral index.  We find that for a given set of model parameter and a
physically motivate $T_0-\gamma$ evolution, the redshift dependence of \GHI
have different shapes for non-equilibrium and equilibrium models.  This
exercise demonstrates the need for accurate determination of thermal history
parameters in order to measure $\Gamma_{\rm HI}(z)$ accurately \citep[also
see][]{becker2013}.  Measuring \GHI accurately is needed to build the physical
models of ionizing sources and UVB.

The non-equilibrium models are computationally more expensive than the corresponding
equilibrium models. 
While constraining $T_0$ and $\gamma$ from observations, one needs to probe a wide range of UVB.
This is usually done by rescaling the photo-heating rates \citep{becker2011,walther2018} and solving equilibrium
ionization evolution. 
We show that one can recover the physically motivated $T_0$ and $\gamma$ in non-equilibrium model from
the equilibrium models generated by rescaling photo-heating rates (Fig. \ref{fig:T0-gamma-recovery}).
Thus one can use equilibrium thermal history models to constrain $T_0$ and $\gamma$ from observations 
provided the \taueffHI in these models match with that from observations and differential
pressure smoothing effects are negligible.


\section*{Acknowledgement}
All the computations are performed using the PERSEUS cluster at IUCAA, Pune and CALX195 machine
at KICC, IoA, Cambridge. 
PG acknowledges the support by the ERC Advanced Grant Emergence-13436. 
We thank the anonymous referee for improving this work and the manuscript.
We also thank Ewald Puchwein for making the non-equilibrium \gthree simulation data 
available for comparison.
We thank Gabor Worseck, George Becker, Laura Keating, Martin Haehnelt, 
Frederick Davies and ENIGMA group for useful discussion.


\bibliographystyle{mnras}
\bibliography{eqbm_non_eqbm} 

\clearpage
\appendix
\section{Details of additional simulations}
\label{app:additional-simulations}

In this section we discuss the details of the additional models analysed in this work.
We run a low resolution simulation with $L=10 \: h^{-1} \: {\rm \cmpc}$ and $N^3_{\rm particle}=128^3$ 
to study the effect of \HI reionization on $T_0$ and $\gamma$ at $z=6$. 
The models presented in \S \ref{sec:simulation} starts from $z=6$ with assumed $T_0$ and $\gamma$. 
In addition, we have developed a semi-numerical method to efficiently compute the thermal history for the 
given UVB from high redshift $z=15$ with better time resolution. 
Such semi-numerical method also provides the insight in to the non-equilibrium effects on TDR.
Furthermore, it is important to check the consistency of our method with self-consistent \gthree simulation.
For this purpose, we use self-consistent equilibrium and non-equilibrium \gthree simulation from \citetalias{puchwein2015}.
The additional models analysed in this work are summarized in Table \ref{tab:additional-model-details}.
In order to distinguish various simulations for different UVB model, we use following nomenclature.
Model L10-N512-G2-EQ-KS19-2.0 refers to $L_{\rm box} = 10 h^{-1} \: $ \cmpc,
$N_{\rm particle} = 512$, \gtwo simulation post-processed with \citecode for equilibrium ionization evolution
using \citetalias{khaire2018a} UVB with SED $\alpha = 1.4$. 
Model L20-N512-G3-NE-HM12-P15 corresponds to $L_{\rm box} = 20 h^{-1} \: $ \cmpc,
$N_{\rm particle} = 512$, \gthree simulation for self-consistent non-equilibrium ionization evolution
using \citetalias{haardt2012} UVB from \citetalias{puchwein2015} paper.

\begin{table*}
\caption{Nomenclature of the models analyse in this work.}
\scalebox{0.9}{
\begin{threeparttable}
\centering
\begin{tabular}{ccccccccc}
\hline \hline
Model Name & $L_{\rm box}$\tnote{a} & $N_{\rm particle}$\tnote{a} & $m_{\rm gas}$\tnote{a} & Code\tnote{b} & Ionization evolution & UVB             & Output Redshift                 & Reference \\
\hline \hline
L10-N512-G2-EQ-HM12                   & 10 & 512          & $10^5$ & \gtwo + \citecode  & equilibrium      & \citetalias{haardt2012}   & $z = 6 - 1.6, \Delta z =0.1$    & This work \\ 
L10-N512-G2-NE-HM12                   & 10 & 512          & $10^5$ & \gtwo + \citecode  & Non-equilibrium  & \citetalias{haardt2012}   & $z = 6 - 1.6, \Delta z =0.1$    & This work \\ 
L10-N512-G2-EQ-KS19-$\alpha$\tnote{d} & 10 & 512          & $10^5$ & \gtwo + \citecode  & equilibrium      & \citetalias{khaire2018a}  & $z = 6 - 1.6, \Delta z =0.1$    & This work \\ 
L10-N512-G2-NE-KS19-$\alpha$\tnote{d} & 10 & 512          & $10^5$ & \gtwo + \citecode  & Non-equilibrium  & \citetalias{khaire2018a}  & $z = 6 - 1.6, \Delta z =0.1$    & This work \\ 
L10-N128-G2-EQ-HM12                   & 10 & 128\tnote{e} & $10^7$ & \gtwo + \citecode  & equilibrium      & \citetalias{haardt2012}   & $z = 15 - 1.6, \Delta z =0.1$   & This work \\ 
L10-N128-G2-NE-HM12                   & 10 & 128\tnote{e} & $10^7$ & \gtwo + \citecode  & Non-equilibrium  & \citetalias{haardt2012}   & $z = 15 - 1.6, \Delta z =0.1$   & This work \\ 
L10-N128-G2-EQ-KS19-$\alpha$\tnote{d} & 10 & 128\tnote{e} & $10^7$ & \gtwo + \citecode  & equilibrium      & \citetalias{khaire2018a}  & $z = 15 - 1.6, \Delta z =0.1$   & This work \\ 
L10-N128-G2-NE-KS19-$\alpha$\tnote{d} & 10 & 128\tnote{e} & $10^7$ & \gtwo + \citecode  & Non-equilibrium  & \citetalias{khaire2018a}  & $z = 15 - 1.6, \Delta z =0.1$   & This work \\ 
L20-N512-G3-EQ-HM12-P15\tnote{c}      & 20 & 512          & $10^6$ & \gthree            & equilibrium      & \citetalias{haardt2012}   & $z = 6 - 1.6, \Delta z =0.1$    & \citetalias{puchwein2015} \\ 
L20-N512-G3-NE-HM12-P15               & 20 & 512          & $10^6$ & \gthree            & Non-equilibrium  & \citetalias{haardt2012}   & $z = 6 - 1.6, \Delta z =0.1$    & \citetalias{puchwein2015} \\ 
SN-EQ-HM12                            & -  & -   & -      & Semi-Numerical     & equilibrium      & \citetalias{haardt2012}   & $z = 15 - 1.6, \Delta z =0.001$ & This work \\ 
SN-NE-HM12                            & -  & -   & -      & Semi-Numerical     & Non-equilibrium  & \citetalias{haardt2012}   & $z = 15 - 1.6, \Delta z =0.001$ & This work \\ 
\hline \hline
\end{tabular}
\begin{tablenotes}
\item[a] $L_{\rm box}$, $N^3_{\rm particle}$ and $m_{\rm gas}$ is length (in $h^{-1}$ \cmpc), number of particles and gas mass resolution (in $\Msun$) in simulation box respectively. 
\item[b] Post-processing module \citecode is applied on \gtwo output. 
         \citecode provides flexibility to change UVB and perform equilibrium / Non-equilibrium ionization evolution efficiently. 
\item[c] We use \citetalias{puchwein2015} simulations to check the consistency of our method. 
\item[d] SED ($\alpha$) is a free parameter in \citetalias{khaire2018a} UVB. 
         We vary $\alpha$ from 1.4 to 2.0 in steps of 0.1 and solve equilibrium / Non-equilibrium ionization evolution equation. 
         However in this work, we present the results only for $\alpha = 1.4, 1.6, 1.8$ and $2.0$.
\item[e] We use low resolution simulation $N_{\rm particle} = 128$ to study the thermal history from $z=15$ to $z=1.6$. 
\end{tablenotes}
\end{threeparttable}
}
\label{tab:additional-model-details}
\end{table*}


\section{Semi-numerical model for thermal history}
\label{app:semi-numerical}
\InputFig{TDR_Analytical.pdf}{75}{%
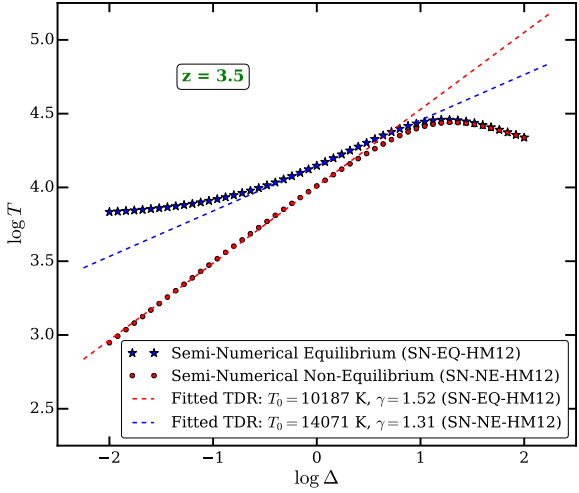
    TDR from semi-numerical approximation at $z=3.5$. 
    TDR deviates from power law at $\log \Delta < -1$ for non-equilibrium model (blue stars).
    TDR in the $\log \Delta$ range -0.5 to 0.5 is fitted with power law and is shown by
    dashed lines (blue, red line for non-equilibrium, equilibrium models respectively).
}{\label{fig:SN-EQ-NE-TDR}}

\InputFig{fHeII_vs_Delta.pdf}{75}{%
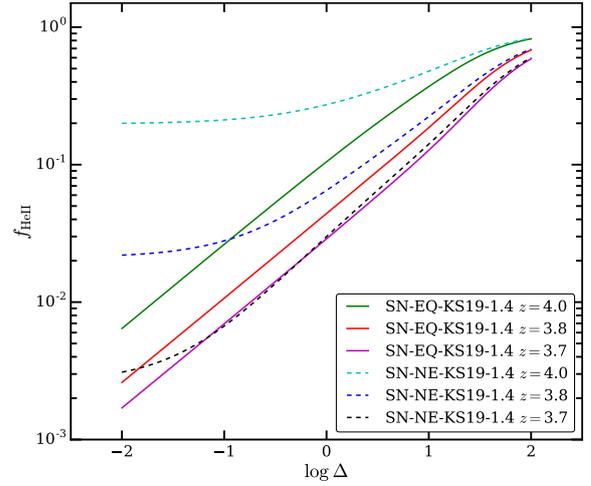
    \fHeII as a function of $\log \Delta$ for SN-EQ-KS19-14 
    and SN-NE-KS19-1.4 models at $z=4.0,3.8,3.7$.
    At all redshifts, \fHeII is proportional to $\Delta^{0.6}$ (Eq. \ref{eq:fHeII-Delta-relation}) for equilibrium models.
    Whereas for non-equilibrium \fHeII is independent of $\Delta$ for $\log \Delta < -0.5$ at
    $z=4.0,3.8$. 
    At $z=3.7$, \fHeII from non-equilibrium model is in good agreement with that from equilibrium model
    and follows $f_{\rm HeII} \propto \Delta^{0.6}$.
}{\label{fig:SN-EQ-NE-fHeII-Delta}}

In this section we present a semi-numerical method (refer as SN-EQ-HM12 and SN-NE-HM12 in Table \ref{tab:additional-model-details}) 
to efficiently predict the evolution of thermal history parameters for a given UVB model. 
This method is fast and we can quickly check the
effect of UVB on thermal history parameters without running full hydro simulation.
This is especially helpful in exploring the parameter space for constraining 
$T_0$ and $\gamma$ from observations (see \S \ref{subsec:T0-gamma-recovery}).
In semi-numerical method, we can start from relatively high redshift and evolve the
temperature of the IGM with better time resolution.
The method is similar to single-cell approximation as described in \citet{puchwein2018}. 
However, we solve the temperature evolution equation at range of $\Delta$ instead of 
temperature evolution at $\Delta=1$ \citep[as done in ][]{puchwein2018}.
We fit a power-law TDR as explained in \S \ref{subsec:TDR}.
As a result, in addition to $T_0$ evolution, we can also predict the evolution in $\gamma$.
The main steps involve in the method are as follows,
\begin{itemize}
    \item First we assume that there are $N=200$ cells with overdensity varying from $\log \Delta = -2.0$ to $2.0$ in steps of $0.02$.
    \item We assume that the overdensity does not evolve with redshift i.e., $d \Delta / dt = 0$ for all cells
        and there is no heating of the gas due to virial shocks $dT_{\rm shock}/dt = 0$.
    \item Under these assumptions, we solve ionization evolution (equilibrium and/or non-equilibrium) Eq. \ref{eq:ionization-evolution-equation} for each cell
        and calculate radiative heating and cooling terms ($dQ/dt$ term).
    \item We then calculate the temperature for each cell from Eq. \ref{eq:temperature-evolution}.
\end{itemize}

Fig. \ref{fig:SN-EQ-NE-TDR} shows the TDR for the cells at $z=3.5$ from equilibrium and non-equilibrium ionization evolution.
As expected, the temperature in non-equilibrium case is consistently higher than that from equilibrium case.
The TDR is also flattened in non-equilibrium case due to density independent photo-heating of the gas
(similar to Fig. \ref{fig:TDR}).
The equilibrium and non-equilibrium TDR are fitted with power-law TDR models as shown by dashed lines.
One can also see the deviation from power-law at small densities in non-equilibrium models
(see Fig. \ref{fig:TDR}).
This deviation in TDR for non-equilibrium model occurs at low densities because \fHeII at these
densities is independent of density.

Fig. \ref{fig:SN-EQ-NE-fHeII-Delta} shows \fHeII as a function of $\log \Delta$ 
for SN-EQ-KS19-1.4 and SN-NE-KS19-1.4 models at $z=4.0,3.8$ and $3.7$.
This choice of redshift is motivated by the fact that $\gamma$ is smallest at $z=4.0$,
$T_0$ is highest at $z=3.8$ and \HeII reionization is completed at $z=3.7$ for KS19-1.4 models
(see Fig. \ref{fig:T0-gamma-evolution-KS19} and Table \ref{tab:HeII-reionization-extent} for our definition).

First  we see a well defined power-law relation between \fHeII and $\Delta$ for equilibrium models
at all $z$. 
This is because \fHeII for photo-ionization equilibrium is given by, 
\begin{equation}\label{eq:fHeII-Delta-relation}
\begin{aligned}
    f_{\rm HeII} &= \frac{n_{\rm HeII}}{n_{\rm He}} = \frac{n_{\rm e} \: n_{\rm HeIII} \: \alpha_{\rm HeIII}(T)}{\Gamma_{\rm HeII} \: n_{\rm He}} \\
    f_{\rm HeII} &\propto \Delta \: T^{-0.7} \\
    f_{\rm HeII} &\propto \Delta^{1-0.7(\gamma-1)} \\
    f_{\rm HeII} &\propto \Delta^{0.6}
\end{aligned}
\end{equation}
where in the last expression we assume, $\gamma \sim 1.57$ 
(see Fig. \ref{fig:T0-gamma-evolution-KS19}) for equilibrium case.
One can see from Fig. \ref{fig:SN-EQ-NE-fHeII-Delta} that the \fHeII and $\Delta$
are well related by power law for equilibrium case. 
The power-law fit to \fHeII vs $\Delta$ curve yields the slope of $\sim 0.61$ for the equilibrium case.
Thus one expects to see the power-law relation between \fHeII and $\Delta$ for
equilibrium case.  

On the other hand, \fHeII at $\log \Delta < 0.5$ is flatter (i.e., independent 
of $\log \Delta$) for SN-NE-KS19-1.4 models at $z=4.0$ and $z=3.8$. 
The volume averaged \fHeII in L10-N512 simulation at $z=4.0,3.8$ is 0.25, 0.055 
(for KS19-1.4 non-equilibrium models) respectively.
Thus at these redshifts the \HeII reionization is still not completed.
As given in Table \ref{tab:HeII-reionization-extent}, the \HeII reionization in 
non-equilibrium case is completed around $z=3.7$ (for KS19-1.4 non-equilibrium model).
At $z=3.7$, we see a good match between \fHeII for equilibrium and non-equilibrium model
in Fig. \ref{fig:SN-EQ-NE-fHeII-Delta}. Note that $z=3.7$ is also the redshift
at which photo-ionization equilibrium is achieved in SN-NE-KS19-1.4 model.

There is no scatter in our TDR for semi-numerical method since 
we have neglected the $d\Delta/dt$ and $dT_{\rm shock}/dt$ term in Eq. \ref{eq:temperature-evolution}.
Thus evolution of density and shock heating of gas is reflected as the scatter in TDR 
(see Fig. \ref{fig:TDR}).

Fig. \ref{fig:T0-gamma-evolution-Low-res-HM12} shows the comparison of 
the $T_0$, $\gamma$ evolution from the semi-numerical method with L10-N512-G2 and
L10-N128-G2 simulations for \citetalias{haardt2012} UVB model.
The differences in $T_0$ and $\gamma$ are less than 11 percent for semi-numerical method.

\section{Scatter in \HeII effective optical depth}
\label{app:tau-eff-HeII-scatter}
In this section, we show the dependence of \taueffHeII scatter on various 
thermal and ionization parameters of IGM. 
Assuming fluctuating Gunn-Peterson relation for \HeII 
(where $K$ is constant that depends on cosmology.),
\begin{equation}
\begin{aligned}
    \tau_{\rm HeII} &= K \: \frac{T_0^{-0.7} \: \Delta^{2-0.7(\gamma-1)}}{\Gamma_{\rm HeII}} \\ \\
\frac{\delta \tau_{\rm HeII}}{\tau_{\rm HeII}} &= -0.7 \: \frac{\delta T_0}{T_0} + [2-0.7(\gamma-1)] \: \frac{\delta \Delta}{\Delta} - \frac{\delta \Gamma_{\rm HeII}}{\Gamma_{\rm HeII}} \\ \\
\end{aligned}
\end{equation}
Since we assume a cosmology and uniform UVB, $\delta \Gamma_{\rm HeII} = 0$,
$\delta T_0 \sim 500 {\rm K}$ (from power-law fit to TDR), $\delta T_0 / T_0 \sim 500/10000 \sim 0.05$ 
and $\delta \Delta / \Delta \sim  0.83$ in our simulations. 
Since $\delta \Delta / \Delta >> \delta T_0 / T_0$,
the contribution of temperature fluctuations to \taueffHeII is sub-dominant.
Therefore, we have
\begin{equation}
\begin{aligned}
    \delta \tau_{\rm HeII} &\propto \frac{\Delta^{1-0.7(\gamma-1)} \: \delta \Delta}{\Gamma_{\rm HeII}} \;\;\; .\\ \\
\end{aligned}
\end{equation}
This suggests that 
(i) the scatter in effective optical depth is directly proportional to fluctuations in density and 
(ii) the scatter in effective optical depth is large when $\Gamma_{\rm HeII}$ is small and vice-versa.
Thus, \taueffHeII scatter in our work is large at $z>3$ compared to \citet{worseck2018} because
our $\Gamma_{\rm HeII} < 4 \times 10^{-15} {\rm s}^{-1}$ the constant value used by \citet{worseck2018}.
On the other hand \taueffHeII scatter in our work is small at $z>3$ compared to \citet{puchwein2018} 
because our \GHeII is larger than \citet{puchwein2018}.
Thus \citet{puchwein2018} will produce large scatter in \taueffHeII than observations.
It is important to note that our UVB evolution is physically motivated than the simple constant 
\GHeII as assumed in \citet{worseck2018}. 

\section{Evolution of mean \HeII effective optical depth}
\label{app:mean-tau-eff-HeII-evolution}
In the literature, the scatter in observed \taueffHeII is usually compared with 
mean \taueffHeII obtained from simulation \citep{puchwein2015,worseck2018}.
Fig. \ref{fig:mean-tau-eff-HeII-evolution} shows the evolution of mean 
\taueffHeII along different quasar sightlines that 
are collected from  \citet{heap2000,zheng2004,fechner2006,syphers2014,worseck2016}.
The observed \taueffHeII is shown for individual quasar sightlines but the observation
sample is not homogenized \citep[exception being][]{worseck2016}.
The red and green shaded regions in the left and right panels respectively show the range 
in \taueffHeII evolution for $1.4 < \alpha < 2.0$.
The observed \taueffHeII scatter is well within the $\alpha=1.4$ and $\alpha=2.0$
region irrespective of equilibrium and non-equilibrium models.
Note the Fig. \ref{fig:tau-eff-HeII-evolution} shows median \taueffHeII evolution with SNR,
redshift path length and instrumental broadening consistent with observations from \citet{worseck2018}.
However, in Fig. \ref{fig:mean-tau-eff-HeII-evolution}, we show the evolution of 
mean \taueffHeII calculated assuming a redshift path length $\Delta z = 0.01$, 
SNR $\sim 10$ and HST-COS LSF.

\InputFigCombine{Mean_HeII_Tau_eff_Evolution.pdf}{150}{%
    Each panel shows the evolution of \textit{mean} \HeII effective optical depth. 
    The observed \taueffHeII from \citet[][green circles]{heap2000}, \citet[][cyan diamonds]{zheng2004},
    \citet[][black squares]{fechner2006}, \citet[][blue stars]{syphers2014} and \citet[][magenta triangles]{worseck2016} 
    are shown in left and right panels. 
    The observed \taueffHeII is shown for individual sightlines but the observation
    sample is not homogenized \citep[exception being][]{worseck2016}.
    Left panel show the \taueffHeII evolution from equilibrium models for $\alpha = 1.4$ (cyan solid line)
    and $\alpha = 2.0$ (blue dashed lines) in \citetalias{khaire2018a} UVB model.
    The red shaded region shows the variation in \taueffHeII due to uncertainty in $\alpha$ for equilibrium models.
    \taueffHeII evolution for $1.4 < \alpha < 2.0$ lies within red shaded region.
    Right panel is similar to left panel except that the \taueffHeII evolution is shown for 
    non-equilibrium models ($\alpha = 1.4$ by black dotted line and $\alpha = 2.0$ by brown dash-dotted line).
    The green shaded region shows the variation in \taueffHeII due to variation in $\alpha$ ($1.4 \leq \alpha \leq 2.0$) for non-equilibrium models.
    Comparison of \taueffHeII evolution from NE-KS19-2.0 with that from EQ-KS19-2.0 
    suggests that the evolution in \taueffHeII is steeper in non-equilibrium models.
    This trend is well correlated with evolution in \fHeII shown in left panel of 
    Fig. \ref{fig:ionization-fraction-combine}.
    Note the Fig. \ref{fig:tau-eff-HeII-evolution} shows median \taueffHeII evolution with SNR,
    redshift path length and instrumental broadening consistent with \citet{worseck2018}.
    In this figure, we show the evolution of mean \taueffHeII calculated assuming a redshift path length
    $\Delta z = 0.01$, SNR $\sim 10$ and LSF of HST-COS.
}{\label{fig:mean-tau-eff-HeII-evolution}}

\section{Uncertainty in evolution of thermal history parameters}
\label{app:uncertainty-in-T0-gamma-evolution}
In this section we discuss the uncertainty in evolution of $T_0$ and $\gamma$ in the models
L10-N512-G2-NE-KS19-1.4 (model NE-KS19-1.4 in main text of the paper).
We run \citecode for L10-N512-G2 simulation from $z=6$ to $z=1.6$ due to limited storage and computational
resources. 
At initial redshift ($z=6$ in L10-N512-G2 case), we assume a power-law TDR for $T_0$ and $\gamma$
consistent with \gthree.
It is important to study the effect of initial $T_0$ and $\gamma$ at $z=6$ on $T_0$ and $\gamma$
redshift at later epoch.
We perform a low resolution L10-N128-G2 simulation to study the effect of initial $T_0$ and $\gamma$.
We store the output for L10-N512-G2 simulation from $z=15$ to $z=1.6$ in steps of $\Delta z = 0.1$.
We post-process the output using \citecode, fit TDR at each redshift and calculate $T_0$, $\gamma$ evolution.
Fig. \ref{fig:T0-gamma-evolution-Low-res-HM12} shows the $T_0$, $\gamma$ evolution from L10-N512-G2 and 
L10-N128-G2 models for \citetalias{haardt2012} equilibrium and non-equilibrium UVB cases.
The $T_0$ and $\gamma$ evolution from the two models match within 8 percent.

Fig. \ref{fig:T0-gamma-evolution-Low-res-KS19} shows the $T_0$,$\gamma$ evolution for
L10-N128-G2 simulation for \citetalias{khaire2018a} equilibrium and non-equilibrium UVB cases.
The $T_0$ and $\gamma$ evolution from $z=11$ to $z=6$ is very similar
for equilibrium and non-equilibrium models.
Furthermore, the $T_0$ and $\gamma$ evolution from $z=15$ to $z=6$
is very similar for different $\alpha$ values.
This is expected as quasar contribution to UVB is not significant at high redshift.
Thus the uncertainty in $T_0$, $\gamma$ (at $z=6$) due to differences in UVB is not significant.

In Fig. \ref{fig:T0-gamma-evolution-Low-res-initial-condition}, we use measurements of $T_0$ 
at $z=6$ from observations of \citet{bolton2012}.
We vary initial $T_0$ from 5000 to 9000 K in \citecode and calculate $T_0$,$\gamma$ evolution.
Fig. \ref{fig:T0-gamma-evolution-Low-res-KS19} shows that even though the $\delta T_0 = 4000$ K at
$z=6$ corresponds to lower $\delta T_0 \sim 500$ K (maximum) at later redshift. 
Since the temperature has an uncertainty of the order of 5 percent, the corresponding 
uncertainty in \GHI would be $\sim 3.5$ percent.

In Fig. \ref{fig:T0-gamma-evolution-observation}, we show the comparison of 
$T_0-\gamma$ measurements from observations with that from equilibrium and non-equilibrium
models for $\alpha = 1.8$ and $2.0$.
The two models produce the \taueffHeII evolution consistent with that from observations 
(see Fig. \ref{fig:tau-eff-HeII-evolution}).
The $T_0$ evolution from \citet{walther2018} is in agreement within $1.5 \sigma$ with L10-N512-NE-KS19-1.8 model.
However, $\gamma$ evolution from \citet{walther2018} is consistent within 2$\sigma$ with that from 
L10-N512-EQ-KS19-1.8 model.

\InputFigCombine{Thermal_history-Plot-0.pdf}{135}{
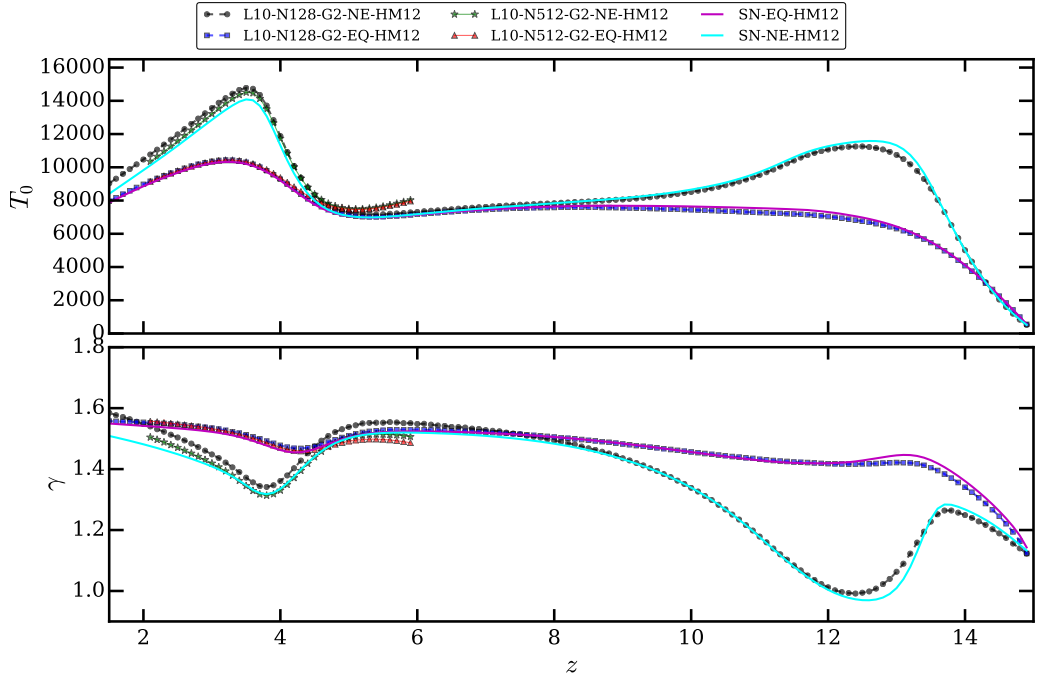
    Same as Fig. \ref{fig:T0-gamma-evolution-KS19} except the $T_0$ and $\gamma$ evolution is shown for 
    L10-N512-G2, L10-N128-G2 and SN (semi-numerical) simulations. 
    In all cases, we use \citetalias{haardt2012} UVB and 
    show $T_0$, $\gamma$ evolution for equilibrium and non-equilibrium ionization evolution.
    For L10-N128-G2-NE-HM12, L10-N128-G2-EQ-HM12, SN-EQ-HM12 and SN-NE-HM12 models we run \citecode from $z=15$ to $z=1.6$.
    While for L10-N512-G2-NE-HM12 and L10-N512-G2-EQ-HM12 models we run \citecode from $z=6$ to $z=2$.
    Figure shows that $T_0$, $\gamma$ evolution in low resolution (L10-N128-G2) simulation is consistent 
    with high resolution (L10-N512-G2) simulation (differences less than 8 percent).
    Whereas the differences in high resolution simulation and semi-numerical method are less 11 percent.
    We use the low resolution simulation to study the effect of \HI reionization on $T_0$ and $\gamma$ values at
    $z=6$.
}{\label{fig:T0-gamma-evolution-Low-res-HM12}}

\InputFigCombine{Thermal_history-Plot-2.pdf}{135}{
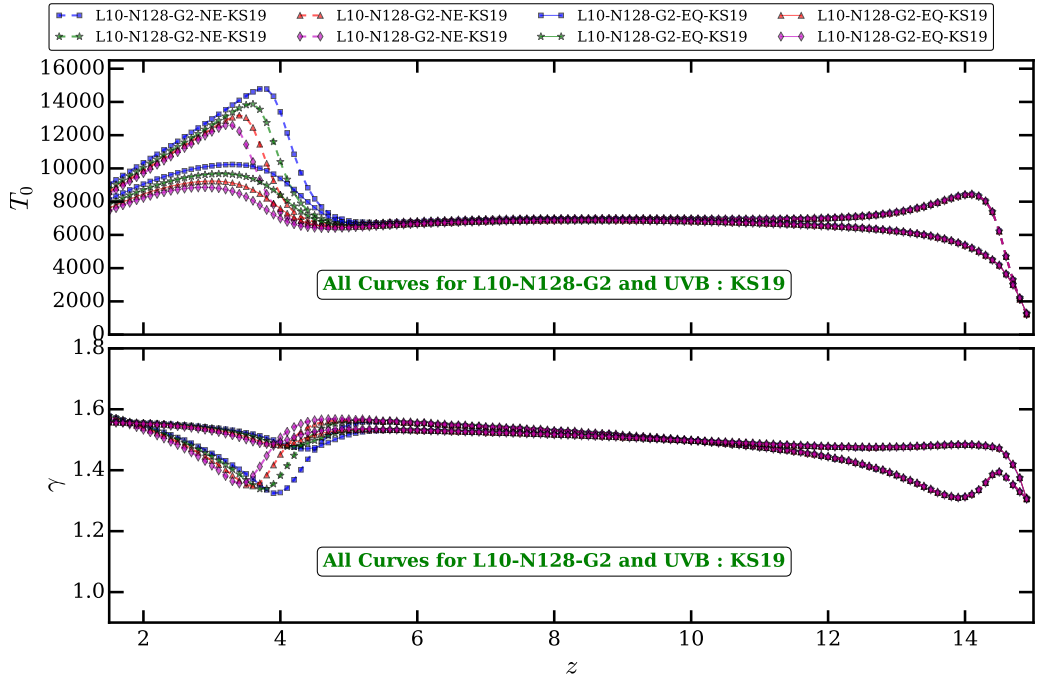
    Same as Fig. \ref{fig:T0-gamma-evolution-KS19} except the $T_0$ and $\gamma$ evolution is shown for 
    L10-N128-G2-KS19-$\alpha$ simulations with $\alpha$ varying from 1.4 to 2.0. 
    The $T_0$ and $\gamma$ values at $z=6$ are quite similar in all models indicating quasar SED has mild effect on $T_0$ 
    and $\gamma$ evolution during \HI reionization.
    The $T_0$ and $\gamma$ values at $z=6$ in \citetalias{khaire2018a} UVB model are less by 6.25 and 6.5 percent than
    that from \citetalias{haardt2012} UVB (not shown in figure) model respectively
}{\label{fig:T0-gamma-evolution-Low-res-KS19}}

\InputFigCombine{Thermal_history-Plot-3.pdf}{150}{
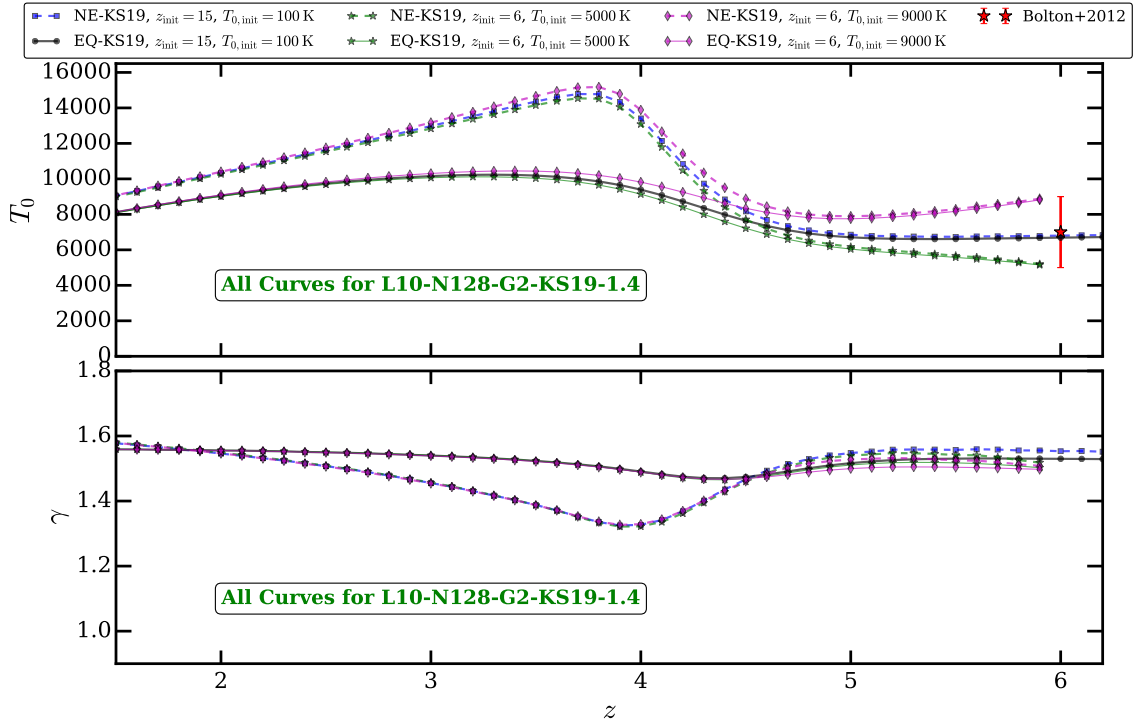
    Same as Fig. \ref{fig:T0-gamma-evolution-KS19} except the $T_0$ and $\gamma$ evolution is shown for 
    L10-N128-G2-KS19-1.4 simulations with variation in initial $T_0$ at $z=6$. 
    We chose initial $T_0 = 5000$ and $9000$ K consistent with \citet{bolton2012} $T_0$ measurement.
    Even though initial $T_0$ varied by $4000$ K, the corresponding maximum variation in $T_0$ is less than 500 K
    at $z<4.5$.
}{\label{fig:T0-gamma-evolution-Low-res-initial-condition}}

\InputFigCombine{Thermal_history_Two_SEDs_Literature.pdf}{150}{
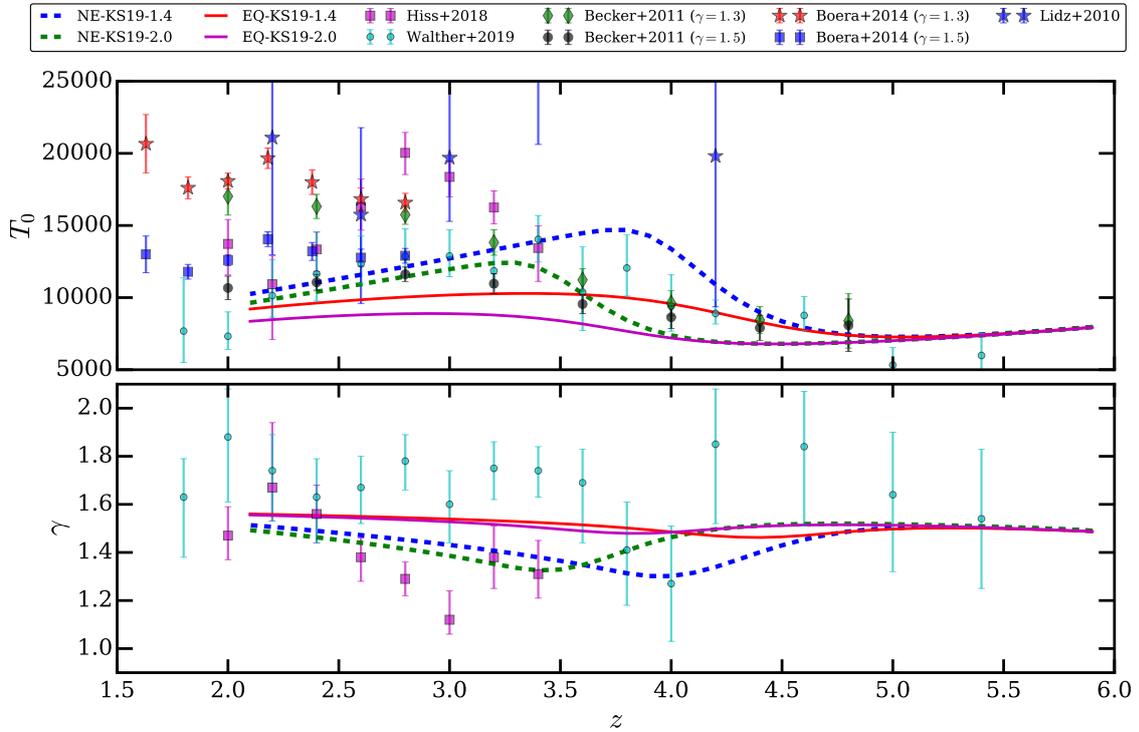
    Comparison of $T_0$ and $\gamma$ evolution from KS19-1.8 and KS19-2.0 models
    with that from observations \citep{lidz2010,becker2011,boera2014,hiss2017,walther2018}. 
    The two models are shown for L10N512 simulation box and produce the \taueffHeII evolution
    consistent with that from observations (see Fig. \ref{fig:tau-eff-HeII-evolution}).
    The $T_0$ evolution from \citet{walther2018} is in agreement (within $1.5 \sigma$) with NE-KS19-1.8 model.
    However, $\gamma$ evolution from \citet{walther2018} is consistent within 2$\sigma$ with that from 
    EQ-KS19-1.8 model.
}{\label{fig:T0-gamma-evolution-observation}}

\section{\HI \lya flux statistics}
\label{app:lya-statistics}
\InputFigCombine{New_Combine_FPDF_FPS_CS_WT.pdf}{180}{%
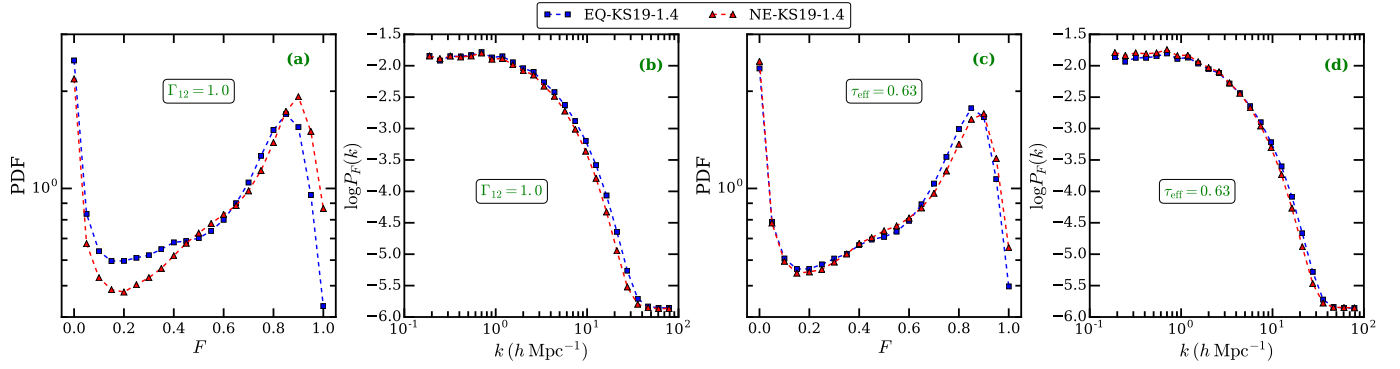
   Panels a,b show the FPDF, FPS respectively from 
   equilibrium and non-equilibrium \citetalias{khaire2018a} UVB models ($\alpha=1.4$) assuming constant \GTW$=1$. 
   \taueffHI is different for equilibrium and non-equilibrium models in this case.
   Panel c and d are similar to panel a and b respectively except that
   the \taueffHI$=0.63$ is same for equilibrium and non-equilibrium models i.e., \GTW is different.
   Comparison of panel a and c suggests that non-equilibrium ionization evolution affect the \GHI measurements. 
   However, the shape of the FPDF is not affected much.
   Panel b and d show that the non-equilibrium model has less power as compared to that for equilibrium
   model at scales $k > 10 \: h \: {\rm Mpc}^{-1}$ due to more pressure smoothing 
   in non-equilibrium model ($T_0$ is larger in non-equilibrium models).
   All plots are shown for $z=3.6$.
}{\label{fig:flux-statistics}}

In this section, we discuss the effect of non-equilibrium ionization evolution on \HI \lya flux
statistics namely flux probability distribution function (FPDF) and flux power spectrum (FPS).
The FPDF and FPS statistics are used in the past to constrain the \GHI.
Whereas FPS is sensitive to $T_0$ and $\gamma$.
This is because for a given cosmology and ignoring the effect of thermal broadening and peculiar velocities, 
the \lya optical optical depth is given by \citep{weinberg1997},
\begin{equation}{\label{eq:optical-depth-approximation}}
    \tau_{\rm HI} \propto \frac{T_0^{-0.7} \: \Delta^{2-0.7(\gamma-1)}}{\Gamma_{\rm HI}}
\end{equation}
where, $\Delta$ is overdensity and we assume that the \HII recombination rate scales
with temperature as $T^{-0.7}$.
The above expression is an approximation and we do not use the for calculation 
of \lya optical depth in our simulated spectra.

The details on how to derive FPDF and FPS from \HI \lya forest spectra are discussed
in \citetalias{gaikwad2018}. Here we briefly show the results for these statistics. 
Fig. \ref{fig:flux-statistics} shows the two flux statistics for EQ-KS19-1.4 and NE-KS19-1.4 models.
Panel a, b show the four statistics assuming a constant \HI photo-ionization rate (\GTW = 1) for two models.
In this case, \taueffHI for the two model is different.
The bottom panels show the four statistics derived for EQ-KS19-1.4 and NE-KS19-1.4 models 
assuming \taueffHI = 0.63.
In this case, we vary \GHI to match the equilibrium and non-equilibrium model \taueffHI with observed 
\taueffHI = 0.63 at $z=3.6$ \citep{becker2013}.
Panel a shows the case where \taueffHI is different for EQ-KS19-1.4 and NE-KS19-1.4 models. 
The FPDF is quite different in two models.
However when we match \taueffHI (by changing \GTW), the FPDF match very well as shown in panel c.
This suggests that non-equilibrium / equilibrium ionization evolution does not affect the shape of the FPDF appreciably.
The shape of the FPDF is relatively insensitive to $T_0$ and $\gamma$ parameters.
Since the thermal history is significantly different for equilibrium and non-equilibrium model 
(see Fig. \ref{fig:T0-gamma-evolution-KS19}), one expects to see the differences in FPS.
On smaller scales ($k > 10 \: h \: {\rm Mpc}^{-1}$), the power in non-equilibrium model is smaller than that for equilibrium model. 
This is because of pressure smoothing is more in non-equilibrium model (since $T_0$ is larger) as compared to equilibrium model.
We again see that the match between EQ-KS19-1.4 and NE-KS19-1.4 model is relatively good when \taueffHI between two models is
same (compare panel b and d in Fig. \ref{fig:flux-statistics}).
Thus non-equilibrium effects during \HeII reionization does not appreciably change the shapes FPDF and FPS derived from \HI \lya forest.
However, the equilibrium and non-equilibrium processes affect the thermal history of IGM.





\bsp	
\label{lastpage}
\end{document}